\documentclass[preprint]{aastex631}
\pdfoutput=1

\usepackage{amsmath}
\usepackage{amssymb}
\usepackage{natbib}
\usepackage{color,soul}
\usepackage{lipsum} 
\usepackage{wasysym}

\submitjournal{PSJ}
\received{January 25}
\revised{March 28}
\accepted{May 3}

\shorttitle{Free Core Nutation and Spin-Over Mode}
\shortauthors{J. Rekier}

\begin{document}

\title{Free Core Nutation and its relation to the Spin-Over Mode}

\correspondingauthor{J\'er\'emy Rekier}
\email{jeremy.rekier@observatory.be}

\author[0000-0003-3151-6969]{J. Rekier}
\affiliation{Royal Observatory of Belgium \\
Avenue circulaire, 3 \\
B-1180 Brussels, Belgium}

\begin{abstract}
The time-varying response of the Earth's and other planets' rotation to external gravitational torques depends strongly on their internal structure. In particular, the existence of the mode known as the Free Core Nutation in the fluid core, is known to amplify the forced nutations in the near-diurnal retrograde frequency band (as measured in the planetary frame of reference). {Due to their proximity in shape and frequency, t}his mode is sometimes equated with {the so-called} Spin-Over Mode {which denotes the free oscillation of a steadily rotating ellipsoidal fluid core}. Through a careful study of the freely rotating two-layer planetary model with a rigid mantle and an inviscid fluid core, {we show that the Spin-Over Mode frequency corresponds to that where} the sum of the external and internal torques on the mantle are balanced, causing {it} to rotate steadily. {The presence of dissipation at the Core-Mantle Boundary causes the Free Core Nutation to become damped and slightly offset its resonance frequency. We show that this offset, which is $\approx-1$ day for the Earth, can be interpreted as the result of the proximity of the Free Core Nutation frequency to that of the Spin-Over Mode, which now corresponds to a minimum in the magnitude of the transfer function for nutations.} {We also show how this proximity leads to a slightly lower Quality factor for the resonance than that compute from the usual formula.}
We conclude {by discussing possible} implications of this mechanism for Mars, the Moon, {and the long-term evolution of the Earth}.
\end{abstract}

\section{Introduction} 
\label{sec:intro}

The orientation of the Earth in space varies in time under the influence of gravitational attraction, primarily by the Sun and Moon, resulting in the motion of its rotation axis known as \textit{precession-nutation}. In the frequency domain, nutation is composed of a wide array of terms -- collectively referred to by the plural \textit{nutations} -- with small amplitude and high frequency compared to the larger, slower precession. The nutations amplitudes depend strongly on the planet's internal structure. In particular, the Earth's liquid core is known to amplify nutations in the retrograde near-diurnal frequency band, as measured in the terrestrial reference frame (see later). This amplification is due to the existence of the free rotational mode known as the \textit{Free Core Nutation} (FCN) inside the fluid core, and coupled to the rotation of the Earth's solid mantle via the pressure torque it exerts on the oblate Core-Mantle Boundary (CMB) \citep{DehantMathews2015}. Such a liquid core is not unique to the Earth but most likely exists in all known terrestrial planets, as well as in the Moon, which must therefore also possess an FCN. 

Planets liquid cores are an ideal subject to the study of rotating fluids given the importance of the \textit{Coriolis force} in these objects. {One example} being the way this force combines with the convective motion in the Earth's core to generate its self-sustaining magnetic field in a complex process known as the \textit{geodynamo}. Already at the linear level, the Coriolis force can act as the restoring force generating oscillations known as \textit{inertial waves} in planetary fluid cores \citep{Greenspan1968}. From a mathematical perspective, it was recently shown that these waves form a complete flow basis in inviscid fluids enclosed within an ellipsoidal CMB, thereby motivating the term \textit{inertial modes}, which is also employed for viscous fluids, by extension \citep{Ivers2015,Backus2017,Ivers2017}.

From laboratory experiments, we know that inertial modes can be excited in many different ways \citep{LeBars2015}. Among existing mechanisms, precession of the container's orientation is particularly interesting given its similarities with the Earth's precession-nutation. Since the work of \citet{Poincare1910}, such forcing is known to excite a simple inertial mode characterized by its near-diurnal frequency and its uniform vorticity parallel to the planet's equatorial plane, when observed from the terrestrial reference frame, later baptized the \textit{Spin-Over Mode} (SOM). Because this mode shares its two defining characteristics with the FCN, the two have sometimes been identified as one and the same thing. To make matters worse, the similarity in name with yet another mode known as the \textit{Tilt-Over Mode} (TOM), has sometimes led to the three terms being used interchangeably \citep[see e.g.][or recently \citealt{NobiliEtAl2021}]{Toomre1974,Noir2003,Cebron2010} causing some confusion between the fluid dynamics, and geodesists communities.

Some attempts to clarify this use of vocabulary have been made recently, starting with the work of \citet{Triana2019} who studied the dynamics of viscous inertial modes coupled to planetary rotation, and showed how the FCN coincides with the SOM, only in the limit where the planet rotates steadily around its axis. \citet{Rekier2020} gave a simplified description of the same problem for the inviscid fluid, and formally showed how the FCN is the direct generalization of the SOM for a freely rotating planet, as well as the only inertial mode affected by the non-steady rotation in the inviscid case. That work can be seen, in large part, as an update to \citet{Hough1895} who had already presented many of the same arguments using a somewhat dated formalism. 

In the present {paper}, {we elaborate on these previous works and present a} detailed analysis of the rotating two-layer planetary model subjected to external gravitational forcing, carried out in Sec.~\ref{sec:2layer}. We find that, while the SOM does indeed {disappear from the spectrum in favour of} the FCN for a {freely} rotating planet, its frequency {retains some significance and} corresponds to that where the amplitudes of the forced nutations become zero {when there is no dissipation at the CMB}. We show how this result follows naturally from the usual {understanding} of the SOM. We also give a detailed description of the TOM and the way it differs to the SOM, and FCN, {in an attempt to remove the confusion between these names}.

As the frequencies of the SOM and FCN are {typically} very close to each other, the amplitude of the forced nutations vary from zero to infinity in a very narrow frequency band {in the idealised non-dissipative model}. In Sec.~\ref{sec:Earth} we reintroduce the effects of the mantle's elasticity, as well as the additional {dissipative} couplings at the CMB. {This coupling causes the FCN to become a damped mode, and to increase its frequency (in absolute value), as is well known. The damping itself further contributes to slightly offset the resonance frequency of the nutations compared to the FCN. By considering the representation of the system in terms of its complex transfer function, we show in Sec.~\ref{sec:discussion} that this offset can be interpreted as a consequence of the proximity between the FCN and the SOM frequencies.}  We {evaluate this offset to be of the order of} $\approx-1$ day {in period} for the Earth, and {likely much smaller for} Mars. We conclude with possible implications for the Moon and the Earth's long term evolution.

\section{Two-layer planetary model}
\label{sec:2layer}
\subsection{Equations of motion}
\label{sec:eqmotion}

The \textit{Liouville equations} governing the rotation of a two-layer planet write:
\begin{align}
\frac{d\mathbf{H}_\mathrm{m}}{dt}+\mathbf{\Omega}\times\mathbf{H}_\mathrm{m}&=\mathbf{\Gamma}_\mathrm{m}+\mathbf{\Gamma}_{\mathrm{f}\rightarrow\mathrm{m}}~,\label{eq:Liouvillemantle}\\
\frac{d\mathbf{H}_\mathrm{f}}{dt}+\mathbf{\Omega}\times\mathbf{H}_\mathrm{f}&=\mathbf{\Gamma}_\mathrm{f}+\mathbf{\Gamma}_{\mathrm{m}\rightarrow\mathrm{f}}~,\label{eq:Liouvillecore}\\
\frac{d\mathbf{H}}{dt}+\mathbf{\Omega}\times\mathbf{H}&=\mathbf{\Gamma}~,\label{eq:Liouville}
\end{align}
where $\mathbf{H}_\mathrm{m}$, and $\mathbf{H}_\mathrm{f}$ respectively denote the angular momentum of the mantle and core, which must add up to the total angular momentum of the whole planet: $\mathbf{H}=\mathbf{H}_\mathrm{m}+\mathbf{H}_\mathrm{f}$. Likewise, $\mathbf{\Gamma}_\mathrm{m}$ and $\mathbf{\Gamma}_\mathrm{f}$ respectively denote the torques exerted on the mantle and core by external sources, and must satisfy: $\mathbf{\Gamma}=\mathbf{\Gamma}_\mathrm{m}+\mathbf{\Gamma}_\mathrm{f}$, where $\mathbf{\Gamma}$ denotes the total torque from external sources. Conservation of angular momentum further imposes that Eqs.~\eqref{eq:Liouvillemantle}~\&~\eqref{eq:Liouvillecore} must add up to Eq.~\eqref{eq:Liouville}, so that:
\begin{equation}
\mathbf{\Gamma}_{\mathrm{m}\rightarrow\mathrm{f}}+\mathbf{\Gamma}_{\mathrm{f}\rightarrow\mathrm{m}}=\mathbf{0}~.
\label{eq:torquebalance}
\end{equation}
i.e. the torque exerted by the mantle on the fluid core must balance that exerted by the core on the mantle. In Eqs.~\eqref{eq:Liouvillemantle} to \eqref{eq:Liouville}, $\mathbf{\Omega}$ denotes the angular velocity of the reference frame, with respect to the inertial frame. It is often convenient to work in the so-called \textit{mantle frame} where the mantle appears at rest, and to use a Cartesian coordinate basis, $\{\hat{\mathbf{x}},\hat{\mathbf{y}},\hat{\mathbf{z}}\}$, aligned with the principal axes of inertia of both the core and mantle -- which are assumed identical. For the axisymmetric planet, the tensor of inertia of the whole planet, and its fluid core, respectively read:
\begin{align}
\underline{\mathbf{I}}&=\mathrm{A}~(\hat{\mathbf{x}}\hat{\mathbf{x}}+\hat{\mathbf{y}}\hat{\mathbf{y}}+(1+e)\hat{\mathbf{z}}\hat{\mathbf{z}})~,\label{eq:TensorofInertia}\\
\underline{\mathbf{I}}_\mathrm{f}&=\mathrm{A}_\mathrm{f}(\hat{\mathbf{x}}\hat{\mathbf{x}}+\hat{\mathbf{y}}\hat{\mathbf{y}}+(1+e_\mathrm{f})\hat{\mathbf{z}}\hat{\mathbf{z}})~,\label{eq:TensorofInertiacore}
\end{align}
where expressions of the form, $\hat{\mathbf{x}}\hat{\mathbf{x}}$, denote the direct (dyadic) product of the unit vector, $\hat{\mathbf{x}}$, with itself.
The real quantities $e>0$, and $e_\mathrm{f}>0$, are the \textit{dynamical flattenings} of the whole planet and its fluid core, respectively. The tensors of inertia satisfy: $\underline{\mathbf{I}}=\underline{\mathbf{I}}_\mathrm{f}+\underline{\mathbf{I}}_\mathrm{m}$, where $\underline{\mathbf{I}}_\mathrm{m}$ is the tensor of inertia of the mantle. The latter being solid, its angular momentum can be expressed in terms of its angular velocity, $\mathbf{\Omega}$, via:
\begin{equation}
\mathbf{H}_\mathrm{m}=\underline{\mathbf{I}}_\mathrm{m}\cdot\mathbf{\Omega}~.
\end{equation}
In the study of planetary rotation, one focuses on small oscillations about a steady state chosen along the $z$-axis:
\begin{equation}
\mathbf{\Omega}=\Omega_0\left(\hat{\mathbf{z}}+\mathbf{m}\right)~,
\label{eq:Omegam}
\end{equation}
where $\Omega_0$ is the diurnal \textit{spin rate}. The vector, $\mathbf{m}$, is called the mantle's \textit{wobble}, and satisfies $|\mathbf{m}|\ll1$. For the Earth, its components are typically of the order of $10^{-8}$ to $10^{-6}$. It is sometimes useful to know the mantle's velocity with respect to some intermediate frame rotating steadily at diurnal frequency around the vertical axis, $\hat{\mathbf{Z}}$, chosen along the mean polar axis of the planet in the inertial frame, hereafter denoted as the \textit{Steadily Rotating Frame} (SRF). The dynamical relations between this frame, the mantle frame, and the inertial frame, are given in Appendix~\ref{sec:frames}. The angular velocity of the mantle in the SRF reads:
\begin{align}
\mathbf{\Omega}_\mathrm{srf}&=\mathbf{\Omega}-\Omega_0\hat{\mathbf{Z}}\nonumber\\
&=\Omega_0\left(\hat{\mathbf{z}}-\hat{\mathbf{Z}}\right)+\Omega_0\mathbf{m}~,
\label{eq:OmegaSRF}
\end{align}

In principle, the angular momentum of the fluid core must be computed from the flow velocity. However, it can be shown that -- in the special case of an ellipsoidal \textit{Core-Mantle Boundary} (CMB) -- this velocity can be decomposed as:
\begin{equation}
\mathbf{v}=\boldsymbol{\omega}_\mathrm{f}\times\mathbf{r}+\mathbf{\nabla}\psi~,
\label{eq:v=wxr+gradpsi}
\end{equation}
where $\mathbf{v}$ denotes the flow velocity with respect to the mantle frame, and $\boldsymbol{\omega}_\mathrm{f}$ is the \textit{mean flow rotation}, satisfying: $\mathbf{\nabla}\times\mathbf{v}=2\boldsymbol{\omega}_\mathrm{f}$. By analogy with Eq.~\eqref{eq:Omegam}, and \eqref{eq:OmegaSRF}, we write:
\begin{align}
\boldsymbol{\omega}_\mathrm{f}&=\Omega_0\mathbf{m}_\mathrm{f}~,\label{eq:omegaf}\\
\boldsymbol{\omega}_\mathrm{f|srf}&=\mathbf{\Omega}_\mathrm{srf}+\boldsymbol{\omega}_\mathrm{f}\nonumber\\
&=\Omega_0\left(\hat{\mathbf{z}}-\hat{\mathbf{Z}}\right)+\Omega_0\left(\mathbf{m}+\mathbf{m}_\mathrm{f}\right)~.\label{eq:OmegafSRF}
\end{align}
where $\boldsymbol{\omega}_\mathrm{f|srf}$ denotes the fluid's angular velocity with respect to the SRF, and $|\mathbf{m}_\mathrm{f}|\ll1$. Equation~\eqref{eq:v=wxr+gradpsi} was first given -- in a different form -- by \citet{Poincare1910}, who also showed that the contribution to the second term of this equation to the core angular momentum was of the second order in the core flattening and could therefore be neglected in front of the other term in planetary applications. The angular momentum of the core, and the whole planet may therefore respectively be written as:
\begin{align}
\mathbf{H}_\mathrm{f}\approx\underline{\mathbf{I}}_\mathrm{f}\cdot\left(\mathbf{\Omega}+\boldsymbol{\omega}_\mathrm{f}\right)~,\label{angmomcore}\\
\mathbf{H}\approx\underline{\mathbf{I}}\cdot\mathbf{\Omega}+\underline{\mathbf{I}}_\mathrm{f}\cdot\boldsymbol{\omega}_\mathrm{f}~,\label{angmomtot}
\end{align}
which are valid to order $\mathcal{O}(|\mathrm{m}|e_\mathrm{f})$. 

Of the three Eqs.~\eqref{eq:Liouvillemantle} to \eqref{eq:Liouville}, only two are independent. The traditional choice is to solve Eq.~\eqref{eq:Liouville} together with a modified version of Eq.~\eqref{eq:Liouvillecore} first proposed by \citet{SasaoEtAl1980}, who showed that it could be advantageously replaced by:
\begin{equation}
\frac{d\mathbf{H}_\mathrm{f}}{dt}-\boldsymbol{\omega}_\mathrm{f}\times\mathbf{H}_\mathrm{f}\approx\mathbf{0}~,
\label{eq:Sasao}
\end{equation}
which avoids having to evaluate $\mathbf{\Gamma}_{\mathrm{m}\rightarrow\mathrm{f}}$, explicitly. Equation~\eqref{eq:Sasao} is valid to the same order of approximation as Eqs.~\eqref{angmomcore}~\&~\eqref{angmomtot}, providing that $\mathbf{\Gamma}_{\mathrm{m}\rightarrow\mathrm{f}}$ is limited to the sum of the pressure and gravitational torques exerted by the mantle on the fluid. We will relax this assumption in Sec.~\ref{sec:Kcmb} below.

Finally, the total torque acting on the planet is the sum of the individual contributions from external gravitational forces and reads:
\begin{equation}
\mathbf{\Gamma}=-\int_\mathcal{V}\rho\mathbf{r}\times\mathbf{\nabla}\phi~dV~,
\label{eq:Gammaext}
\end{equation}
where the integral runs over the volume, $\mathcal{V}$, of the whole planet, $\rho$ is the mass density, and $\phi$ is the total gravitational potential. 
For an axisymmetric planet, and to first order in $e$, the two equatorial components of Eq.~\eqref{eq:Gammaext} in the body frame can be combined as \citep{Mathews1991}:
\begin{equation}
\tilde{\Gamma}=i\mathrm{A}e\tilde{\phi}~,
\label{eq:Gammaexttilde}
\end{equation}
where we have defined, $\tilde{\Gamma}\equiv\Gamma^x+i\Gamma^y$, and where $\tilde{\phi}$ is proportional to the coefficient of the degree $2$, tesseral spherical harmonics expansion of $\phi$.

From the above, the equatorial components of the wobble, $\mathrm{m}^x$ and $\mathrm{m}^y$, can be shown to decouple from the axial component $\mathrm{m}^z$, to first order in $\mathbf{m}$, and similarly for the components of $\mathbf{m}_\mathrm{f}$. After solving for these vectors components in the mantle frame, we can use the solution to compute the kinetic energy of the mantle and core. In practice, it will often prove useful to do so in the SRF, where they are defined as:
\begin{align}
E_\mathrm{kin}^\mathrm{m}&=\frac{1}{2}\mathbf{\Omega}_\mathrm{srf}^\intercal\cdot\underline{\mathbf{I}}_\mathrm{m}\cdot\mathbf{\Omega}_\mathrm{srf}~,\label{eq:Ekinm}\\
E_\mathrm{kin}^\mathrm{f}&=\frac{1}{2}\boldsymbol{\omega}_\mathrm{f|srf}^\intercal\cdot\underline{\mathbf{I}}_\mathrm{f}\cdot\boldsymbol{\omega}_\mathrm{f|srf}~.\label{eq:Ekinf}
\end{align}

\subsection{Prograde and retrograde motions (sign conventions)}
\label{sec:prograderetrograde}

The wobbly motion of an axisymmetric planetary model can be written as a circular rotation of the vector, $\mathbf{m}$, within the equatorial plane at constant angular frequency, $\omega$\footnote{Note that this is \textit{not} true for a non-axisymmetric planet \citep[see e.g.][]{DehantMathews2015}.}:
\begin{equation}
\left(
\begin{array}{c}
\mathrm{m}^x\\
\mathrm{m}^y
\end{array}
\right)
=
\underbrace{
\left(
\begin{array}{cc}
\cos\omega t  & -\sin\omega t \\
\sin\omega t  & \cos\omega t \\
\end{array}
\right)}
_{R(t)}
\left(
\begin{array}{c}
\mathrm{m}_0^x\\\mathrm{m}_0^y
\end{array}
\right)~,
\label{eq:mxmy(t)}
\end{equation}
where $\mathrm{m}^x_0$, and $\mathrm{m}^x_0$ are the values of $\mathrm{m}^x$, and $\mathrm{m}^y$ at $t=0$. The \textit{canonical basis} offers a useful way to diagonalise the relation Eq.~\eqref{eq:mxmy(t)}. It is defined in terms of the cartesian basis as:
\begin{equation}
\left(
\begin{array}{c}
\hat{\boldsymbol{\epsilon}}_-\\
\hat{\boldsymbol{\epsilon}}_+
\end{array}
\right)
=
\underbrace{
\frac{1}{\sqrt{2}}
\left(
\begin{array}{cc}
1 & -i \\
-1 & -i \\
\end{array}
\right)}
_{P}
\left(
\begin{array}{c}
\hat{\mathbf{x}}\\
\hat{\mathbf{y}}
\end{array}
\right)~.
\end{equation}
It is then straightforward to show that the equatorial components of $\mathbf{m}$ along $\hat{\boldsymbol{\epsilon}}_-$, and $\hat{\boldsymbol{\epsilon}}_+$, satisfy:
\begin{equation}
\left(
\begin{array}{c}
\mathrm{m}^-\\
\mathrm{m}^+
\end{array}
\right)
=P^*R(t)P^\intercal
\left(
\begin{array}{c}
\mathrm{m}^-_0\\
\mathrm{m}^+_0
\end{array}
\right)~.
\end{equation}
where $*$ denotes the complex conjugate. Upon defining $\tilde{\mathrm{m}}\equiv(\mathrm{m}^x+i\mathrm{m}^y)=\sqrt{2}~\mathrm{m}^-$, we finally obtain:
\begin{equation}
\left(
\begin{array}{c}
\tilde{\mathrm{m}}\\
\tilde{\mathrm{m}}^*
\end{array}
\right)=
\left(
\begin{array}{cc}
\mathrm{e}^{i\omega t} & 0 \\
0 & \mathrm{e}^{-i\omega t} \\
\end{array}
\right)
\left(
\begin{array}{c}
\tilde{\mathrm{m}}_0\\
\tilde{\mathrm{m}}^*_0
\end{array}
\right)~.
\label{eq:m-m+(t)}
\end{equation}
From Eq.~\eqref{eq:m-m+(t)}, we see that $\tilde{\mathrm{m}}$, and $\tilde{\mathrm{m}}^*$, rotate in opposite directions in the complex plane. For $\omega>0$, $\tilde{\mathrm{m}}$ describes a \textit{prograde} motion, and $\tilde{\mathrm{m}}^*$ describes a \textit{retrograde} motion with the same angular velocity. The opposite is true when $\omega<0$. We also see that the dynamics of $\tilde{\mathrm{m}}$, and $\tilde{\mathrm{m}}^*$ decouple for a circular motion. This allows us to focus exclusively on $\tilde{\mathrm{m}}$ in what follows. However, based on the above, it should be kept in mind that, in order for a general solution Eq.~\eqref{eq:mxmy(t)} to be real-valued, to each particular solution, $\tilde{\mathrm{m}}_0(\omega)\mathrm{e}^{i\omega t}$, there must correspond another particular solution with opposite frequency: $\tilde{\mathrm{m}}_0^*(-\omega)\mathrm{e}^{-i\omega t}$. Both solutions describe the same physical prograde, or retrograde motion.

In what follows, we drop the subscript notation for readability, and use the same symbol, $\tilde{\mathrm{m}}$, to denote both the wobble component and its \textit{Fourier mode}, $\tilde{\mathrm{m}}_0(\omega)$.

\subsection{Solution for a rigid mantle}
\label{sec:twolayerrigid}

{In the rest of this work, we use} units in which the (prograde) diurnal frequency has the numerical value: $\Omega_0=1$, and all frequencies are given in \textit{cycles per day} (cpd), unless otherwise mentioned. Based on the above considerations, Eqs.~\eqref{eq:Liouville} \& \eqref{eq:Sasao} can be summarized in the following matrix form:
\begin{equation}
\left(
\begin{array}{cc}
 \omega -e  & (1+\omega)\frac{\mathrm{A}_\mathrm{f}}{\mathrm{A}}\\
 \omega  & \omega+\left(1+e_\mathrm{f}\right) \\
\end{array}
\right)
\left(
\begin{array}{c}
 \tilde{\mathrm{m}} \\
 \tilde{\mathrm{m}}_\mathrm{f}
\end{array}
\right)
=
\left(
\begin{array}{c}
 e\tilde{\phi}\\
 0
\end{array}
\right)~,
\label{eq:twolayermatrix}
\end{equation}
where $\tilde{\mathrm{m}}_\mathrm{f}\equiv\left(\mathrm{m}_\mathrm{f}^x+i\mathrm{m}_\mathrm{f}^y\right)$ is analogous to $\tilde{\mathrm{m}}$ for the fluid core. 

Equation~\eqref{eq:twolayermatrix} reflects an important property already identified by \citet{Poincare1910} who noticed that, when $\omega=-1$, the dynamics of $\tilde{\mathrm{m}}$, becomes independent of $\tilde{\mathrm{m}}_\mathrm{f}$, so that the wobble of the two-layer model is governed by the same equation as for a single-layer \textit{rigid} planet:
\begin{equation}
(\omega-e)\tilde{\mathrm{m}}=e\tilde{\phi}~.
\label{eq:singlelayerrigideq}
\end{equation}
\citeauthor{Poincare1910} named this phenomenon \textit{gyrostatic rigidity} and showed how it is related to the existence of a particular type of rotational mode which was later called the \textit{Tilt-Over Mode} (TOM) (see Sec.~\ref{sec:SOM} below). The solution to Eq.~\eqref{eq:singlelayerrigideq}, writes:
\begin{equation}
\tilde{\mathrm{m}}_\mathrm{R}(\omega)=\frac{e\tilde{\phi}}{(\omega-e)}~.
\label{eq:mtildesinglelayerrigid}
\end{equation}
where the subscript is a reminder that Eq.~\eqref{eq:mtildesinglelayerrigid} is only valid for a single-layer rigid planet for $\omega\neq-1$. This solution becomes infinite at the frequency:
\begin{equation}
\omega_\mathrm{E}=e~,
\label{eq:omegaE}
\end{equation}
which corresponds to the resonance with the well-known \textit{Eulerian Free Wobble} of rigid-body dynamics \citep[see e.g.][]{Landau1969}. Equation~\eqref{eq:mtildesinglelayerrigid} also corresponds to the solution of Eq.~\eqref{eq:twolayermatrix} in the limit of $\tilde{\mathrm{m}}_\mathrm{f}=0$. This is expected, as in this case the liquid core rotates uniformly with the mantle and the whole planet behaves as a solid body. The other limit $\tilde{\mathrm{m}}=0$ corresponds to the situation where the planet rotates steadily around the $z$-axis. In this case, we find that the liquid core must oscillate with frequency:
\begin{equation}
\omega_\mathrm{som}=-(1+e_\mathrm{f})~.
\label{eq:omegasom}
\end{equation}
This is nothing other than the frequency of the SOM of a fluid inside a \textit{steadily}-rotating (oblate) ellipsoid. This frequency is not usually written as in Eq.~\eqref{eq:omegasom} {\citep[see however][]{Smith1977}}. In Appendix \ref{sec:fluiddyn} we show that this expression is equivalent to the more usual formula derived from the momentum conservation equation of rotating fluid dynamics. 

Using the definitions of Eqs.~\eqref{eq:omegaE} \& \eqref{eq:omegasom}, we may rewrite Eq.~\eqref{eq:twolayermatrix} as:
\begin{equation}
\left(
\begin{array}{cc}
\omega -\omega_\mathrm{E}  & (1+\omega)\frac{\mathrm{A}_\mathrm{f}}{\mathrm{A}} \\
 \omega  & \omega-\omega_\mathrm{som}\\
\end{array}
\right)
\left(
\begin{array}{c}
 \tilde{\mathrm{m}} \\
 \tilde{\mathrm{m}}_\mathrm{f}
\end{array}
\right)
=
\left(
\begin{array}{c}
 e\tilde{\phi}\\
 0
\end{array}
\right)~.
\label{eq:matrixeulersom}
\end{equation}
We find the free rotational modes by solving the above for $\tilde{\phi}=0$. There are two independent eigenfrequencies corresponding to the roots of the characteristic polynomial:
\begin{equation}
\Delta\equiv\left(\omega-\omega_\mathrm{E}\right)\left(\omega-\omega_\mathrm{som}\right)-\omega(1+\omega)\frac{\mathrm{A}_\mathrm{f}}{\mathrm{A}}=0~.
\label{eq:detomega}
\end{equation}
At this point, it should be reminded that the solutions to Eq.~\eqref{eq:detomega} -- based on Eq.~\eqref{eq:Sasao} -- are only expected to be valid to first order in the flattening parameters, $e$, and $e_\mathrm{f}$. To this order, the first solution writes:
\begin{align}
\omega_\mathrm{fcn}&=-1+\frac{\mathrm{A}}{\mathrm{A}_\mathrm{m}}(1+\omega_\mathrm{som})\label{eq:omegaFCNSOM}\\
&=-1-\frac{\mathrm{A}}{\mathrm{A}_\mathrm{m}}e_\mathrm{f}~,\label{eq:omegaFCN}
\end{align}
which we recognise as the frequency of the \textit{Free Core Nutation} (FCN), also-known as the \textit{Nearly Diurnal Free Wobble} (NDFW) due to its proximity with the (retrograde) diurnal frequency in the mantle frame. This proximity makes it a very important mode in the study of nutations as it tends to be greatly amplified by the TOM, a phenomenon that \citet{Poincare1910} described as a `double resonance'. The second solution writes:
\begin{align}
\omega_\mathrm{cw}&=\frac{\mathrm{A}}{\mathrm{A}_\mathrm{m}}\omega_\mathrm{E}\label{eq:omegaCWE}\\
&=\frac{\mathrm{A}}{\mathrm{A}_\mathrm{m}}e~,\label{eq:omegaCW}
\end{align}
which corresponds to the frequency of the mode known as the \textit{Chandler Wobble} (CW), named after the first astronomer to measure its frequency for the Earth. This measured frequency is quite different to Eq.~\eqref{eq:omegaCW} because of the combined effects of the mantle's elasticity -- which are considered in Sec.~\ref{sec:Elastic} below -- and surface processes \citep[see e.g.][]{RekierEtAl2021}. 

\begin{figure}
   \centering
   \includegraphics[width=1\textwidth]{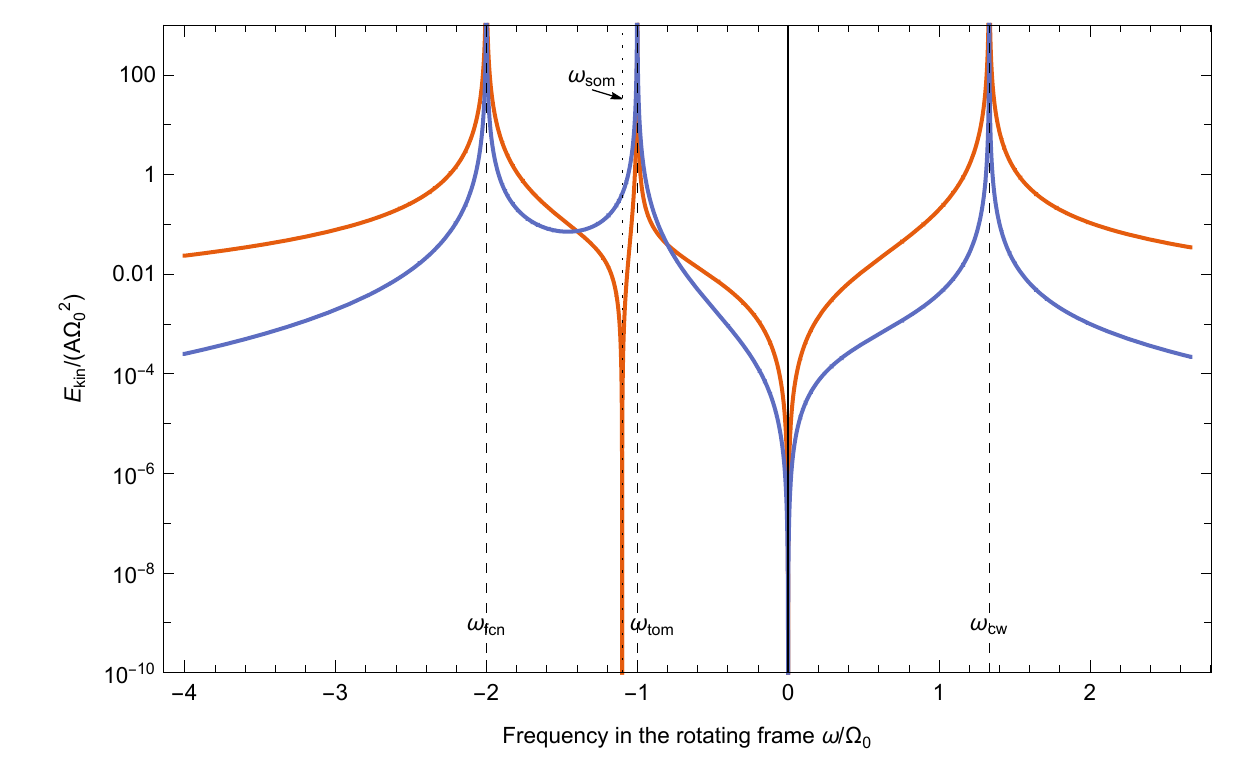} 
   \caption{Normalised kinetic energy of the mantle (in red), and core (in blue) as a function of the forcing frequency ($\mathrm{A}_\mathrm{m}/\mathrm{A}=1/10$, $e_\mathrm{f}=1/10$, $e=2/15$, and $\tilde{\phi}=1$).}
   \label{fig:Ekin_2layer}
\end{figure}

Finally, the general solution to Eq.~\eqref{eq:matrixeulersom} may be written formally as:
\begin{equation}
\left(
\begin{array}{c}
 \tilde{\mathrm{m}} \\ \tilde{\mathrm{m}}_\mathrm{f}
\end{array}
\right)
=\frac{\mathrm{A}}{\mathrm{A}_\mathrm{m}}\frac{e\tilde{\phi}}{\left(\omega-\omega_\mathrm{fcn}\right)\left(\omega-\omega_\mathrm{cw}\right)}
\left(
\begin{array}{c}
\omega-\omega_\mathrm{som} \\ -\omega
\end{array}
\right)~.\label{eq:mmfsolutiontwolayer}
\end{equation}
Interestingly, we see that $\tilde{\mathrm{m}}=0$ when $\omega=\omega_\mathrm{som}$, something that can be also observed also on Fig.~\ref{fig:Ekin_2layer} showing the (normalised) kinetic energy of the mantle (in red) and the core (in blue) as measured in the SRF, for $\mathrm{A}_\mathrm{m}/\mathrm{A}=1/10$, $e_\mathrm{f}=1/10$, and $e=2/15$, corresponding to a planet with a large (heavy) fluid core and a comparatively small (light) mantle.
We see that the kinetic energies of both the mantle, and the core are infinite at the frequencies $\omega_\mathrm{fcn}$, and $\omega_\mathrm{cw}$ indicated by dashed lines, as expected from Eq.~\eqref{eq:mmfsolutiontwolayer}, and that these also become infinite when $\omega=-1$ corresponding to the resonance with the TOM (see the next section). The kinetic energy of the mantle is zero at $\omega_\mathrm{som}$, indicated by the dotted line, whereas that of the core remains finite. {The presence of a third resonance at $\omega=-1$ might seem counter-intuitive. It comes from our choice to plot our results in the SRF, whose orientation with respect to the mantle frame is given by Eq.~\eqref{eq:wobblenutation} which, when added to the system Eq.~\eqref{eq:matrixeulersom} introduces the additional resonance corresponding to the TOM (see next section).}

{Another thing to mention is that although from Eq.~\eqref{eq:mmfsolutiontwolayer} it might seem that $\tilde{\mathrm{m}}$ and $\tilde{\mathrm{m}}$ become infinite at the resonances, this is never the case in practice, as non-linear effects disregarded here starts to be important before it happens. This is also true for the other results given below.}

\subsection{The Spin-Over and Tilt-Over modes}
\label{sec:SOM}
The SOM is a free mode of the two-layer model's fluid core \textit{only when the mantle's rotation is steady} \citep{Triana2019,Rekier2020}. {From Eq.~\eqref{eq:mmfsolutiontwolayer}, we see that its frequency, $\omega_\mathrm{som}$, also corresponds to that where the amplitude of the mantle's motion is zero, $\tilde{\mathrm{m}}=0$}. We can understand the reason {for this} by taking the ratio of the two components of the solution:
\begin{equation}
\tilde{\mathrm{m}}_\mathrm{f}=-\frac{\omega~\tilde{\mathrm{m}}}{\omega-\omega_\mathrm{som}}~.
\label{eq:mforcedmf}
\end{equation}
Equation~\eqref{eq:mforcedmf} can be arrived at by working directly from the momentum equation of fluid dynamics applied to the core flow (see Appendix~\ref{sec:fluiddyn}). This formula can also be obtained directly by considering the second row of Eq.~\eqref{eq:matrixeulersom} which, in isolation, can be interpreted as describing the wobbly response of the liquid core, $\tilde{\mathrm{m}}_\mathrm{f}$, subjected to a \textit{prescribed} wobbly perturbation, $\tilde{\mathrm{m}}$ at the forcing frequency, $\omega$. When $\tilde{\mathrm{m}}$ is constant such perturbation corresponds to a steady precession of the mantle around the $z$-axis. In this case, it is a well-known fact that the flow response has a resonance at the SOM frequency, in agreement with Eq.~\eqref{eq:mforcedmf} \citep{Poincare1910,Busse1968,NoirCebron2014}. 

By contrast, the fluid core wobble of a freely rotating planet remains finite when $\omega=\omega_\mathrm{som}$, as both numerator and denominator in Eq.~\eqref{eq:mforcedmf} become zero. {Looking back at the equatorial projection of Eq.~\eqref{eq:Liouvillemantle} and setting $\tilde{\mathbf{m}}=0$ there, we can see that this corresponds to the frequency where the torque produced on the mantle by the external potential is exactly balanced by that produced by the fluid core,} leaving the mantle in a state of steady rotation. 

The SOM frequency is close, although importantly \textit{not} equal to that of the TOM, even though the two names have sometimes been used interchangeably, {as mentioned in the introduction}. Whereas the SOM owes its existence to that of the planet's fluid core, and its frequency depends on the core's flattening, the TOM exists for \textit{all} rotating objects, regardless of their shape or internal structure, and its frequency is always \textit{exactly} equal to $\omega=-1$. {The TOM has already been described in details by many authors \citep[see e.g.][]{Smith1977,Wahr1981,Rogister2001}}. In Appendix~\ref{sec:frames}, we show that {this frequency is equal to zero, when measured} in the inertial frame, or equivalently {that the associated motion has an} infinite period, $T\rightarrow\infty$. If an external torque with such period were to be applied on the planet {-- such as that produced on the spheroidal Earth's figure by a stationary object at a positive celestial latitude}, it would respond by \textit{tilting} its orientation in space permanently in the direction of {the torque}. When observed in the mantle frame, this reorientation corresponds to an exactly diurnal retrograde motion of the rotation axis around the planet's figure axis. The importance of this mode for the nutation, can be understood by considering the dynamic relation between the 
wobble and the nutation amplitude given at the bottom of Appendix~\ref{sec:frames}.


\subsection{Transfer function}
\label{sec:transferfunction}

An important concept in the study of nutations is that of \textit{transfer function} defined as the ratio of the wobble computed for the planetary model considered to the rigid wobble given by Eq.~\eqref{eq:mtildesinglelayerrigid}:
\begin{equation}
\mathrm{T}(\omega)\equiv\frac{\tilde{\mathrm{m}}(\omega)}{\tilde{\mathrm{m}}_\mathrm{R}(\omega)}=\frac{\tilde{\eta}(\omega+1)}{\tilde{\eta}_\mathrm{R}(\omega+1)}~,
\label{eq:transferfunctiondefinition}
\end{equation}
where $\mathrm{T}$, should not be confused with the period, $T$. The second equality follows from Eq.~\eqref{eq:wobblenutation} of Appendix~\ref{sec:frames} \citep[see also][]{DehantMathews2015}. This function, allows to separate the computation of nutations into two parts. In the first step, the nutations are computed from ephemerides for an hypothetical rigid planet with flattening, $e$. The actual nutations can then be obtained for the more realistic planetary model as a second step by multiplication with $\mathrm{T}(\omega)$ containing all the information on the planet's internal structure.

\begin{figure}
   \centering
   \includegraphics[width=0.8\textwidth]{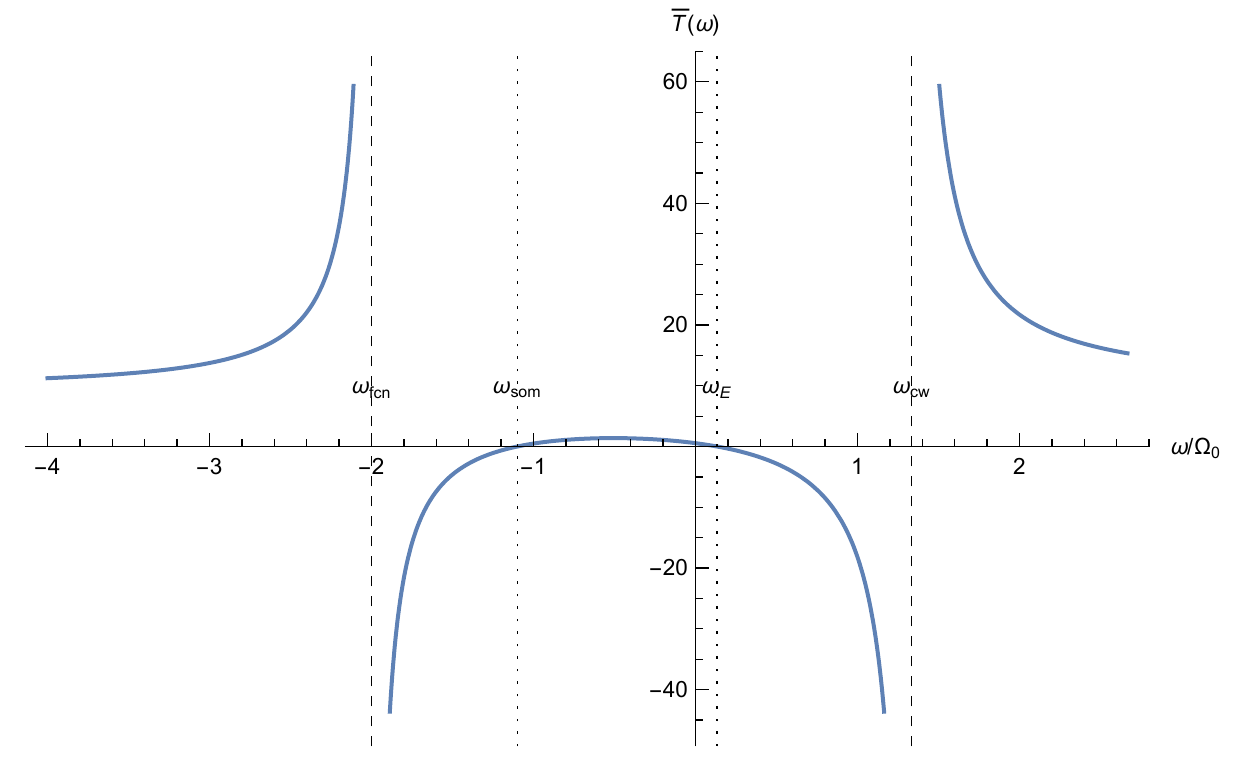} 
   \caption{Normalised transfer function of a (rigid) two-layer planet with a large (heavy) liquid core and a comparatively small (light) mantle ($\mathrm{A}_\mathrm{m}/\mathrm{A}=1/10$, $e_\mathrm{f}=1/10$, and $e=2/15$).}
   \label{fig:transferfunction}
\end{figure}

Using the first component of Eq.~\eqref{eq:mmfsolutiontwolayer}, into Eq.~\eqref{eq:transferfunctiondefinition}, we obtain the expression for the transfer function of the two-layer model:
\begin{equation}
\mathrm{T}(\omega)=\frac{\mathrm{A}}{\mathrm{A}_\mathrm{m}}\frac{\left(\omega-\omega_\mathrm{E}\right)\left(\omega-\omega_\mathrm{som}\right)}{\left(\omega-\omega_\mathrm{cw}\right)\left(\omega-\omega_\mathrm{fcn}\right)}~.
\label{eq:transferfunctiontwolayer}
\end{equation}
{This transfer function does not satisfy the gyrostatic rigidity condition exactly, which imposes that $\mathrm{T}(-1)=1$ (see Sec.~\ref{sec:twolayerrigid}, above). That is because the formalism used to arrive at \eqref{eq:transferfunctiontwolayer} is only valid to first order in the planet's flattening. This can be remedied by introducing the normalised transfer function:
\begin{equation}
\bar{\mathrm{T}}(\omega)=\frac{\mathrm{A}}{\mathrm{A}_\mathrm{m}}\frac{(1+\omega_\mathrm{cw})}{(1+\omega_\mathrm{E})}\frac{\left(\omega-\omega_\mathrm{E}\right)\left(\omega-\omega_\mathrm{som}\right)}{\left(\omega-\omega_\mathrm{cw}\right)\left(\omega-\omega_\mathrm{fcn}\right)}~,
\label{eq:transferfunctiontwolayernormalised}
\end{equation}
which has all the right properties, and also satisfies $\bar{\mathrm{T}}(-1)=1$ exactly. Figure~\ref{fig:transferfunction} shows this normalised transfer function for the same parameters as Fig.~\ref{fig:Ekin_2layer}. Incidentally, note that gyrostatic rigidity prohibits making the approximation, $e(\omega-\omega_\mathrm{som})\approx e(\omega+1)$ \citep[which is the limit taken by e.g.][]{DehantMathews2015a}, in the first component of Eq.~\eqref{eq:mmfsolutiontwolayer}, even though this might seem as the consistent choice to first order in $e$ and $e_\mathrm{f}$. Finally, note that $\mathrm{T}(\omega)$ is close to $\bar{\mathrm{T}}(\omega)$ for all values of $\omega$ providing that: 
\begin{equation}
\omega_\mathrm{cw}-\omega_\mathrm{E}=\frac{\mathrm{A}_\mathrm{f}}{\mathrm{A}_\mathrm{m}}e\ll1~.
\label{eq:conditiongyrostaticrigidity}
\end{equation}
This is true in most planetary applications where $e$ is small, and the inertia of the mantle {is much bigger than the core's}. For example, in the Earth, $\mathrm{A}_\mathrm{f}/\mathrm{A}_\mathrm{m}\approx1/9$ (see Table \ref{tab:parameters}).
}

The expression of the transfer function in Eq.~\eqref{eq:transferfunctiontwolayer} {(or \eqref{eq:transferfunctiontwolayernormalised})} is quite satisfying. Its \textit{zeros}, $\omega_\mathrm{E}$, and $\omega_\mathrm{som}$, respectively correspond to the free modes of oscillations of the mantle and the core taken in isolation, i.e. of the freely rotating mantle shell with a hollow centre (for which $\mathrm{A}_\mathrm{f}=0$) and the steadily rotating fluid core (for which $\mathrm{A}_\mathrm{m}\rightarrow\infty$). On the other hand, its \textit{poles}, $\omega_\mathrm{cw}$, and $\omega_\mathrm{fcn}$, correspond to the two rotational modes of the fully coupled freely rotating two-layer system. 
It should be noted, however, that although the two roots, $\omega_\mathrm{E}$, and $\omega_\mathrm{som}$, are treated on the same footing in Eq.~\eqref{eq:transferfunctiontwolayer}, they have a somewhat different physical interpretation: the former ensures that the amplitude of the non-rigid wobble remains finite when $\omega=\omega_\mathrm{E}$, whereas the latter introduces a `true' zero, as is clearly visible from Fig.~\ref{fig:Ekin_2layer}. This zero can have a detectable effect on the amplitudes of the FCN and nutations amplitudes in the Earth and other planets, as will be discusses in the next section.

\section{Application to the Earth}
\label{sec:Earth}
\subsection{Rigid mantle}

\begin{figure}
   \centering
   \includegraphics[width=1\textwidth]{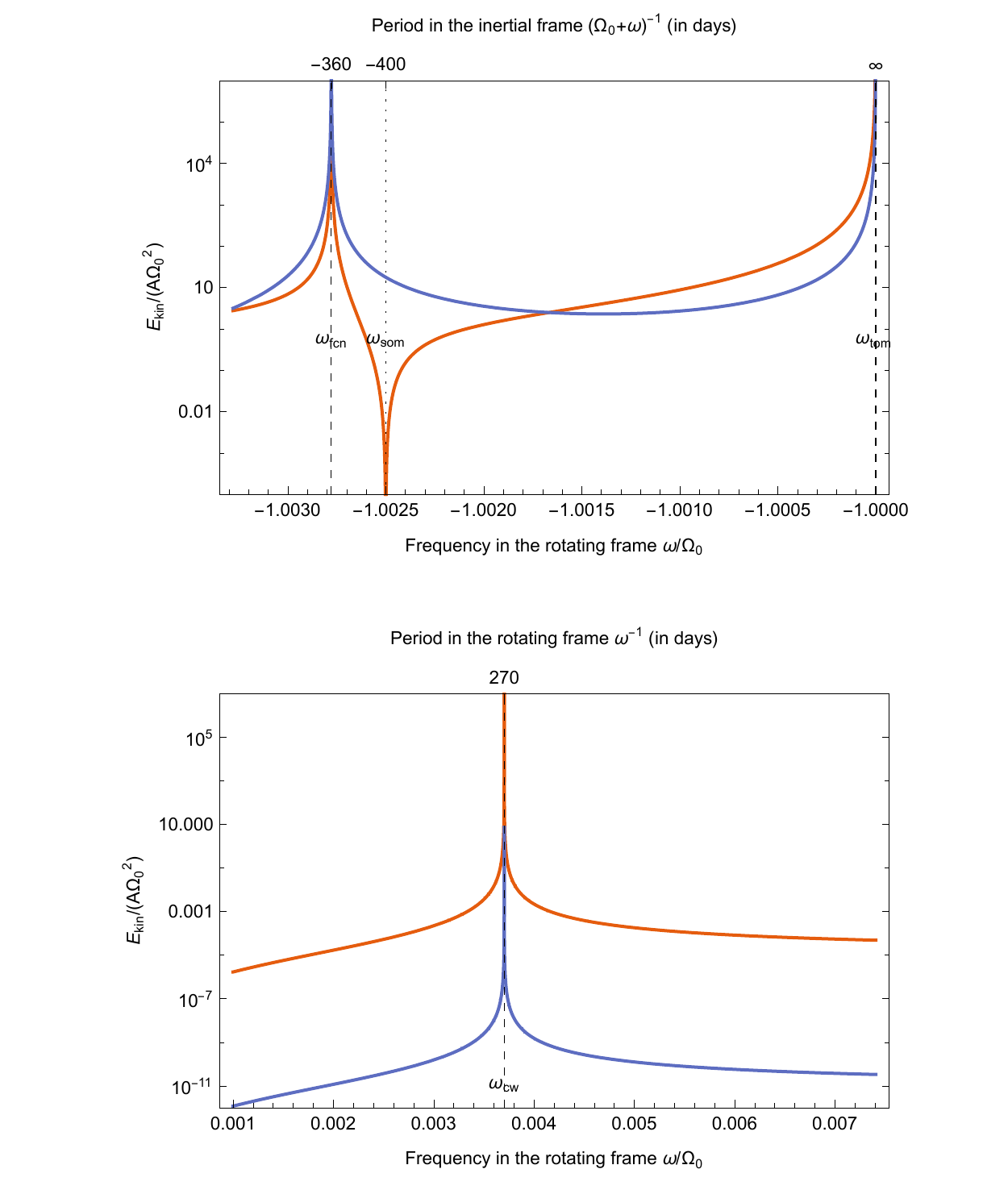} 
   \caption{Normalised kinetic energy of the mantle (in red) and core (in blue) as a function of the forcing frequency ($\mathrm{A}_\mathrm{m}/\mathrm{A}=9/10$, $e_\mathrm{f}=1/400$, $e=1/300$, and $\tilde{\phi}=1$). Top: centred around $\omega_\mathrm{fcn}$. Bottom: centred around $\omega_\mathrm{cw}$.}
   \label{fig:Ekin_2layer_Earth}
\end{figure}

In the example of the previous section, we have used values of the parameters that are quite different to those found in planetary applications. This was done in order to accentuate the differences between $\omega_\mathrm{fcn}$ and $\omega_\mathrm{som}$ on the one hand, and $\omega_\mathrm{cw}$ and $\omega_\mathrm{E}$ on the other. In particular, we had $\mathrm{A}_\mathrm{m}<\mathrm{A}_\mathrm{f}$, whereas, as we have already mentioned, the opposite is true in most cases. Also, the values of $e$ and $e_\mathrm{f}$ for real planets are typically one order of magnitude smaller than those we have used. 
Figure~\ref{fig:Ekin_2layer_Earth} shows the normalised kinetic energy of the mantle (in red) and core (in blue), similarly to Fig.~\ref{fig:Ekin_2layer} for values of the parameters closer to those found in the Earth ($\mathrm{A}_\mathrm{m}/\mathrm{A}=9/10$, $e_\mathrm{f}=1/400$, $e=1/300$, and $\tilde{\phi}=1$). The top figure is centred around $\omega_\mathrm{fcn}$, and the bottom figure around $\omega_\mathrm{cw}$.
The values of $\omega_\mathrm{fcn}$ and $\omega_\mathrm{som}$ are so close in this case that it is easier to compare their periods {(in days)} as measured in the inertial frame of reference based on the following formula \citep[see][and Appendix~\ref{sec:frames}]{Dehant2017}:
\begin{equation}
T_\mathrm{inertial}=(\Omega_0+\omega)^{-1}~.
\end{equation}
Accordingly, the period of the TOM, visible on the right edge of the figure, is infinite. On the other hand, the period of the CW is more readable in the rotating frame (SRF) according to the following formula:
\begin{equation}
T_\mathrm{srf}=\omega^{-1}~.
\end{equation}
One important difference compared to the example of Fig.~\mbox{\ref{fig:Ekin_2layer}}, is that the kinetic energy of the core is larger than that of the mantle in the vicinity of $\omega=\omega_\mathrm{fcn}$ (they are both infinite at that precise frequency). This is consistent with the traditional interpretation of the FCN as a core motion which is reflected in its name. What this study shows, is that this is even more true at $\omega=\omega_\mathrm{som}$ where the mantle is immobile. The opposite situation is true in the vicinity of $\omega=\omega_\mathrm{cw}$, where the kinetic energy of the mantle dominates over that of the core.
The numerical values of $T_\mathrm{fcn}=-360$ days in the inertial frame, and $T_\mathrm{cw}=270$ days in the rotating frame, are quite different to those measured for the real Earth where elasticity plays an important role to lengthen the periods of those modes as we now discuss.

\subsection{Elastic mantle}
\label{sec:Elastic}

Real planets are never perfectly rigid, they deform under the effects of gravity and internal pressure forces. In the case of an hypothetical single-layer planet, elasticity can be shown to shorten the Eulerian frequency to \citep{MunkMacDonald1960}:
\begin{equation}
\bar{\omega}_\mathrm{E}\approx\omega_\mathrm{E}\left(1-\frac{k_2}{k_s}\right)~,\label{eq:EulerelasticLove}
\end{equation}
where $k_2$, and $k_s=3eG\mathrm{A}/(\Omega_0^2R^5)>k_2$, respectively denote the degree-2 elastic, and secular Love numbers, and where $G$, and $R$, are the gravitational constant, and the mean planetary radius. Because of the accidental proximity between Eq.~\eqref{eq:EulerelasticLove}, and the measured CW frequency for the Earth, $\bar{\omega}_\mathrm{E}$ has sometimes been identified with that mode. In reality, the CW frequency of the real Earth is altered by {the oceans present} at its surface {\citep{SmithDahlen1981,Wahr1984a}}, and by the existence of the fluid core {\citep[][and below]{DehantMathews2015}}.

The effects of elasticity on planetary rotation can be incorporated in models through the use of \textit{compliance parameters} measuring the changes in the tensor of inertia of the planet, and its fluid core: $\Delta\underline{\mathbf{I}}$, and $\Delta\underline{\mathbf{I}}_\mathrm{f}$. To leading order, in the wobble, it can be shown that one needs only consider the two combinations: $\tilde{\mathrm{\Delta I}}\equiv\left(\mathrm{\Delta I}^{xz}+i\mathrm{\Delta I}^{yz}\right)$, and $\tilde{\mathrm{\Delta I}}_\mathrm{f}\equiv\left(\mathrm{\Delta I}_\mathrm{f}^{xz}+i\mathrm{\Delta I}_\mathrm{f}^{yz}\right)$. Following \citet{SasaoEtAl1980}, these may then be parametrised as:
\begin{align}
\tilde{\mathrm{\Delta I}}&=\mathrm{A}~\left[\kappa\left(\tilde{\mathrm{m}}-\tilde{\phi}\right)+\xi\tilde{\mathrm{m}}_\mathrm{f}\right]~,\label{eq:DeltaItwolayer}\\
\tilde{\mathrm{\Delta I}}_\mathrm{f}&=\mathrm{A}_\mathrm{f}\left[\gamma\left(\tilde{\mathrm{m}}-\tilde{\phi}\right)+\beta\tilde{\mathrm{m}}_\mathrm{f}\right]~,\label{eq:DeltaIftwolayer}
\end{align}
where $\kappa$, $\xi$, $\gamma$, and $\beta$, are the four compliance parameters.

Remarkably, \citeauthor{SasaoEtAl1980} have shown that their Eq.~\eqref{eq:Sasao} remains valid to the same order of approximation provided that the compliance parameters are of the same order as the dynamical flattenings, $e$ and $e_\mathrm{f}$, which is typically the case in all planetary applications. Upon using Eqs.~\eqref{eq:DeltaItwolayer} \& \eqref{eq:DeltaIftwolayer}, Eq.~\eqref{eq:twolayermatrix} is now replaced by \citep{Mathews1991}:
\begin{equation}
\left(
\begin{array}{cc}
\omega-e+\kappa(1+\omega)  & (1+\omega)\left(\frac{\mathrm{A}_\mathrm{f}}{\mathrm{A}}+\xi\right)\\
(1+\gamma)\omega  & (1+\beta)\omega+\left(1+e_\mathrm{f}\right) \\
\end{array}
\right)
\left(
\begin{array}{c}
 \tilde{\mathrm{m}} \\
 \tilde{\mathrm{m}}_\mathrm{f}
\end{array}
\right)
=
\tilde{\phi}
\left(
\begin{array}{c}
e-\kappa(1+\omega)\\
 -\gamma\omega
\end{array}
\right)~.
\label{eq:twolayermatrixelastic}
\end{equation}
Importantly, by setting $\omega=-1$ into Eq.~\eqref{eq:twolayermatrixelastic}, we see that the elastic two-layer system satisfies the gyrostatic rigidity constraint.

By analogy with the Sec.~\ref{sec:twolayerrigid}, we define the SOM of the steadily rotating two-layer elastic planet as the solution of Eq.~\eqref{eq:twolayermatrixelastic} for $\tilde{\mathrm{m}}=\tilde{\mathrm{\phi}}=0$. This gives the frequency: 
\begin{align}
\bar{\omega}_\mathrm{som}&=-\frac{1+e_\mathrm{f}}{1+\beta}\label{eq:SOMelastic}\\
&\approx\omega_\mathrm{som}+\beta~.\label{eq:SOMelasticapprox}
\end{align}
Since {$\beta>0$}, and $\omega_\mathrm{som}<0$, we have $|\bar{\omega}_\mathrm{som}|<|\omega_\mathrm{som}|$. As expected, we see that elasticity shortens the SOM frequency, similarly to the Eulerian wobble.

The free modes frequencies of the coupled system are the roots of the characteristic polynomials of the matrix in Eq.~\eqref{eq:twolayermatrixelastic}, which upon using the above definitions for $\bar{\omega}_\mathrm{E}$ and $\bar{\omega}_\mathrm{som}$, may be written as:
\begin{equation}
\Delta=(\omega-\bar{\omega}_\mathrm{E})(\omega-\bar{\omega}_\mathrm{som})-\frac{(1+\gamma)}{(1+\beta)(1+\kappa)}\left(\frac{\mathrm{A}_\mathrm{f}}{\mathrm{A}}+\xi\right)\omega(1+\omega)=0~.
\end{equation}
To first order in the dynamical flattening and compliance parameters, the two solutions corresponding to the `elastic' FCN and CW, are respectively \citep[e.g.][]{Mathews1991,MathewsEtAl2002}:
\begin{align}
\bar{\omega}_\mathrm{fcn}&=-1-\frac{\mathrm{A}}{\mathrm{A}_\mathrm{m}}(e_\mathrm{f}-\beta)~,\label{eq:omegafcnelastic}\\
\bar{\omega}_\mathrm{cw}&=\frac{\mathrm{A}}{\mathrm{A}_\mathrm{m}}(e-\kappa)~.\label{eq:omegacwelastic}
\end{align}
Finally, to the same order, the solution to Eq.~\eqref{eq:twolayermatrixelastic} can be written as:
\begin{align}
\tilde{\mathrm{m}}&=\frac{\mathrm{A}}{\mathrm{A}_\mathrm{m}}\frac{\left(e-(1+\omega)\kappa+\epsilon\omega\right)}{\left(\omega-\bar{\omega}_\mathrm{fcn}\right)\left(\omega-\bar{\omega}_\mathrm{cw}\right)}\left(\omega-\frac{e\bar{\omega}_\mathrm{som}+\epsilon}{e-\epsilon}\right)\tilde{\phi}~,\label{eq:mtwolayerelastic}\\
\tilde{\mathrm{m}}_\mathrm{f}&=-\frac{\mathrm{A}}{\mathrm{A}_\mathrm{m}}\frac{\left(e-(1+\omega)\kappa+\gamma\omega\right)}{\left(\omega-\bar{\omega}_\mathrm{fcn}\right)\left(\omega-\bar{\omega}_\mathrm{cw}\right)}\omega\tilde{\phi}~,
\label{eq:mftwolayerelastic}
\end{align}
where we have introduced the following new parameter: 
\begin{equation}
\epsilon=\frac{\mathrm{A}_\mathrm{f}}{\mathrm{A}}\gamma~,
\end{equation}
which satisfies $0<\epsilon<\gamma$. We see that, contrary to {what we might have guessed from} the rigid case, $\tilde{\mathrm{m}}$ is not zero at $\omega=\bar{\omega}_\mathrm{som}$, but rather at another frequency equal to:
\begin{equation}
\bar{\omega}_\mathrm{som}^\epsilon\equiv\frac{e\bar{\omega}_\mathrm{som}+\epsilon}{e-\epsilon}\lessapprox\bar{\omega}_\mathrm{som}~.
\label{eq:omegaSOMepsilon}
\end{equation}
We can understand this result by realising that {any given flow inside the core} produces a torque on the {elastic} mantle {thereby changing} its moments of inertia. Eq.~\eqref{eq:omegaSOMepsilon} gives the forcing frequency at which this change exactly balances the additional momentum. When the moment of inertia of the mantle is much bigger than the fluid's, corresponding to the limit $\mathrm{A}_\mathrm{f}/\mathrm{A}\rightarrow0$, then $\epsilon\rightarrow0$, and $\bar{\omega}_\mathrm{som}^\epsilon\rightarrow\bar{\omega}_\mathrm{som}$. {The variable $\omega_\mathrm{som}^\epsilon$---where $\epsilon$ should not be confused with an exponent---is chosen to acknowledge its relation with the SOM.}

Elasticity also introduces two new roots in $\tilde{\mathrm{m}}$, and $\tilde{\mathrm{m}}_\mathrm{f}$, respectively at: $\omega=(e-\kappa)/(\kappa-\epsilon)$, and $\omega=(e-\kappa)/(\kappa-\gamma)$. These correspond to the frequencies at which the total torque acting respectively on the whole planet, or on the fluid core is exactly balanced by a change in their inertia tensors. Something that is not possible for a rigid mantle.

\begin{figure}
   \centering
   \includegraphics[width=1\textwidth]{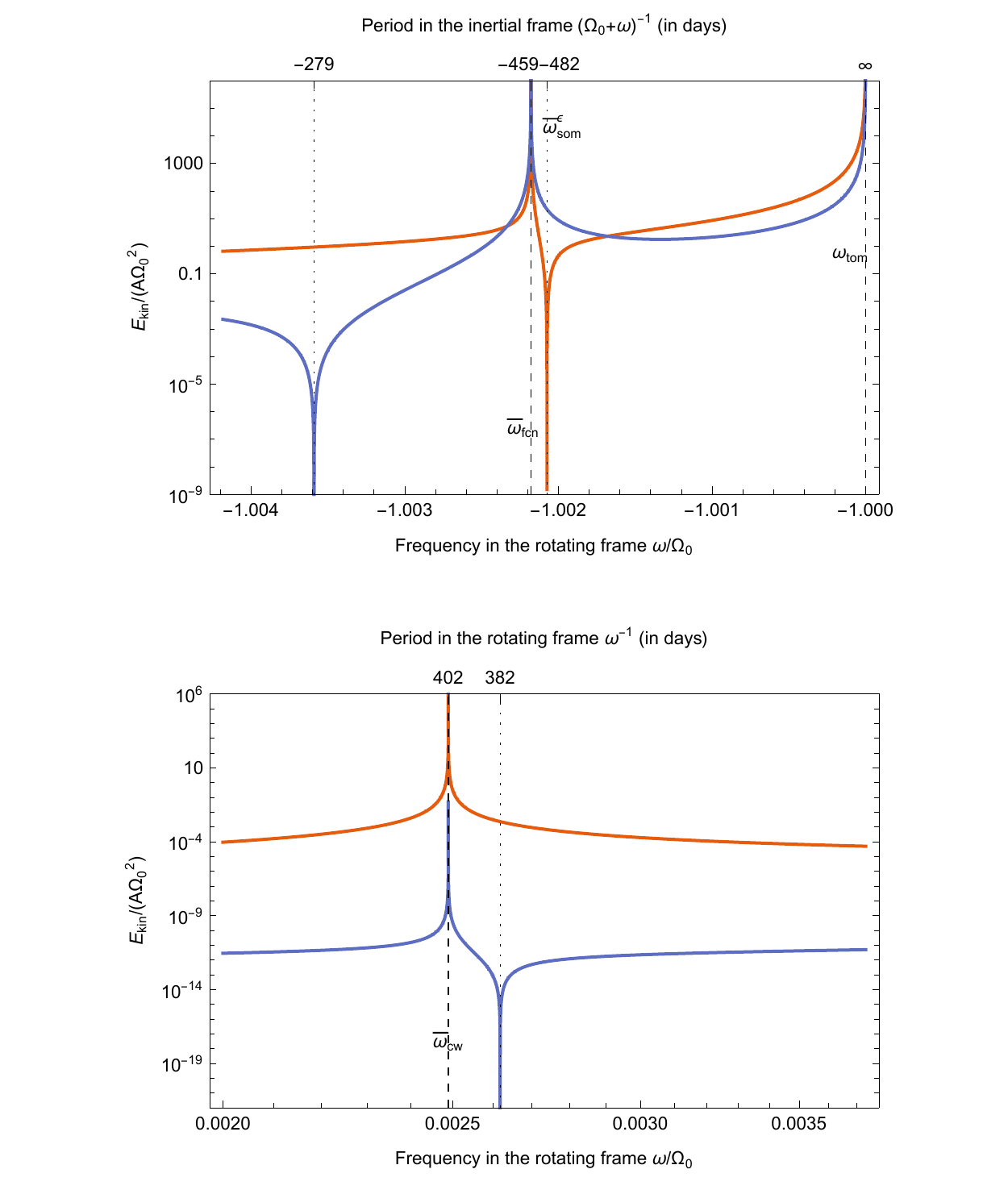} 
   \caption{Normalised kinetic energy of the mantle (in red) and core (in blue) for the two-layer elastic planet as a function of the forcing frequency ($\tilde{\phi}=1$, other parameters are given in Table~\ref{tab:parameters}). Top: centred around $\bar{\omega}_\mathrm{fcn}$. Bottom: centred around $\bar{\omega}_\mathrm{cw}$.}
   \label{fig:Ekin_2layer_Earth_elastic}
\end{figure}
Figure~\ref{fig:Ekin_2layer_Earth_elastic} shows the normalised kinetic energy of the mantle (in red) and fluid core (in blue) in the SRF for an elastic two-layer planet resembling the Earth, with values of the parameters given in Table~\ref{tab:parameters}. 
\begin{table}
\begin{center}
\begin{tabular}{c|ccccccccc}
Symbol & $\mathrm{A}_\mathrm{m}/\mathrm{A}$ & $e\times10^3$ & $e_\mathrm{f}\times10^3$ & $\kappa\times10^3$ & $\xi\times10^3$ & $\beta\times10^3$ & $\gamma\times10^3$ & $K_\mathrm{cmb}\times10^3$\\
\hline
Earth & $0.887$ & $3.285$ & $2.548$ & $1.039$ & $0.222$ & $0.616$ & $1.965$ & $0.128 - i0.019$
\end{tabular}
\end{center}
\caption{Ratio of the equatorial moments of inertia, dynamical flattenings, compliance parameters, and CMB coupling constant for the Earth \citep{DehantMathews2015,ZhuEtAl2017}.}
\label{tab:parameters}
\end{table}
A somewhat surprising feature of this figure is the fact that there is not just one, but two frequencies at which the kinetic energy of the fluid core is zero (indicated by dotted lines), and neither of these correspond to the value of $\omega=(e-\kappa)/(\kappa-\gamma)$ derived above. The reason for this is that the total kinetic energy of the fluid core in the SRF is proportional to the square of the sum of the angular velocities of the core and mantle (see Eq.~\ref{eq:OmegafSRF}). The kinetic energy of the core is therefore zero when those two vectors add up to zero. {The expressions of the two values of the forcing frequency, $\omega_0$, where this balance takes place are quite long and not very informative. However, they can be shown to} satisfy:
\begin{equation}
\tilde{\mathrm{m}}_\mathrm{f}(\omega_0)=-\frac{\omega_0\tilde{\mathrm{m}}(\omega_0)}{\omega_0+1}~.
\label{eq:Ekinmf0}
\end{equation}
This circumstance is unique to the elastic two-layer model and does not exist in the rigid model where $\tilde{\mathrm{m}}$ and $\tilde{\mathrm{m}}$ satisfy Eq.~\eqref{eq:mforcedmf} for all values of $\omega$, which is incompatible with Eq.~\eqref{eq:Ekinmf0}, except in the trivial (spherical) limit where $\omega_\mathrm{som}\rightarrow-1$. Note that, for the parameters of Table~\ref{tab:parameters}, the difference between the values of $\bar{\omega}_\mathrm{som}$, and $\bar{\omega}_\mathrm{som}^\epsilon$---though small---corresponds to a non-negigible difference of period of $\approx-35$ days measured in the inertial frame, bringing $\bar{\omega}_\mathrm{som}^\epsilon$ closer to the value of $\bar{\omega}_\mathrm{fcn}$.

\subsection{Additional coupling at the CMB}
\label{sec:Kcmb}

In addition to the gravitational and pressure torques which are accounted for in Eq.~\eqref{eq:Sasao}, there are other torques at the CMB contributing to the exchange of angular momentum between the core and mantle. It is customary to parametrise those extra torques by introducing a single parameter, $\mathbf{\Gamma}_\mathrm{cmb}$, into the right-hand side of Eq.~\eqref{eq:Sasao}:
\begin{equation}
\frac{d\mathbf{H}_\mathrm{f}}{dt}-\boldsymbol{\omega}_\mathrm{f}\times\mathbf{H}_\mathrm{f}=\mathbf{\Gamma}_\mathrm{cmb}~.
\label{eq:SasaoGammaCMB}
\end{equation}
This most notably includes the \textit{electromagnetic torque} produced by the induced electric currents at the base of the mantle {\citep{Buffett1992,BuffettEtAl2002,Dumberry2012}}, combined with the \textit{viscous torque} caused by the frictions between the mantle and the viscous fluid core {\citep{MathewsGuo2005,DeleplaceCardin2006,koot2010}}. The computation of the exact dependence of $\mathbf{\Gamma}_\mathrm{cmb}$ on those two physical phenomena and others is highly non-trivial. For simplicity, it is customary to use the following parametrisation attributed to \citet{SasaoEtAl1980}:
\begin{equation}
\mathbf{\Gamma}_\mathrm{cmb}=-\Omega_0\mathrm{A}_\mathrm{f}\left(K_1\hat{\mathbf{z}}\times(\boldsymbol{\omega}_\mathrm{f}-\boldsymbol{\omega}_\mathrm{f}^\parallel)-K_2(\boldsymbol{\omega}_\mathrm{f}-\boldsymbol{\omega}_\mathrm{f}^\parallel)-K_3\boldsymbol{\omega}_\mathrm{f}^\parallel\right)~,
\label{eq:Gammacmb}
\end{equation}
where $\boldsymbol{\omega}_\mathrm{f}^\parallel\equiv(\boldsymbol{\omega}_\mathrm{f}\cdot\hat{\mathbf{z}})\hat{\mathbf{z}}$ is the axial part of $\boldsymbol{\omega}_\mathrm{f}$, and where $K_1$, $K_2$, and $K_3$ are real-valued \textit{coupling constants}. {Eq.~\eqref{eq:Gammacmb} is similar in shape to the formulae given by \citet{Noir2003} and \citet{NoirCebron2014} based on the work of \citet{Busse1968} for the viscous torque of a fluid on an ellipsoidal boundary.} The first term in {this equation} accounts for the torque exerted by the mantle on the fluid core by viscous, and electromagnetic -- and possibly other -- forces. The second term accounts for the \textit{power dissipated} by those forces. The third term of Eq.~\eqref{eq:Gammacmb} only affects the planet's spin rate, and has no effect on the wobble, as can be seen by combining the two equatorial components of $\mathbf{\Gamma}_\mathrm{cmb}$ in the usual fashion to yield:
\begin{equation}
\tilde{\Gamma}_\mathrm{cmb}=-iK_\mathrm{cmb}\Omega_0^2\mathrm{A}_\mathrm{f}\tilde{\mathrm{m}}_\mathrm{f}~.
\label{eq:Gammatildecmb}
\end{equation}
where we have defined:
\begin{equation}
K_\mathrm{cmb}\equiv K_1+iK_2~.
\end{equation}
The values of $K_1$, and $K_2$ are typically of the same order of magnitude as $e$, $e_\mathrm{f}$, and the compliance parameters in planetary applications (see Table~\ref{tab:parameters}). We should emphasise that the torque given by Eq.~\eqref{eq:Gammacmb} is \textit{not} equivalent to having $\mathbf{\Gamma}_\mathrm{cmb}\sim K_\mathrm{cmb}\boldsymbol{\omega}_\mathrm{f}$, contrary to what one might guess by working directly from Eq.~\eqref{eq:Gammatildecmb}.

Introducing Eq.~\eqref{eq:Gammatildecmb} into Eq.~\eqref{eq:SasaoGammaCMB}, the dynamical equations for the elastic two-layer planet then become:
\begin{equation}
\left(
\begin{array}{cc}
\omega-e+\kappa(1+\omega)  & (1+\omega)\left(\frac{\mathrm{A}_\mathrm{f}}{\mathrm{A}}+\xi\right)\\
(1+\gamma)\omega  & (1+\beta)\omega+\left(1+e_\mathrm{f}+K_\mathrm{cmb}\right) \\
\end{array}
\right)
\left(
\begin{array}{c}
 \tilde{\mathrm{m}} \\
 \tilde{\mathrm{m}}_\mathrm{f}
\end{array}
\right)
=
\tilde{\phi}
\left(
\begin{array}{c}
e-\kappa(1+\omega)\\
 -\gamma\omega
\end{array}
\right)~.
\label{eq:twolayermatrixelasticKcmb}
\end{equation}
Notice that $K_\mathrm{cmb}$ only appears in the second row of the matrix. By inspection of Eq.~\eqref{eq:twolayermatrixelasticKcmb}, we immediately obtain the frequency of the SOM in the coupled case:
\begin{align}
\bar{\omega}_\mathrm{som}&=-\frac{1+e_\mathrm{f}+K_\mathrm{cmb}}{1+\beta}\nonumber\\
&\approx\omega_\mathrm{som}+\beta-K_\mathrm{cmb}~.\label{eq:SOMelasticKcmb}
\end{align}
Similarly, the frequency of the FCN now reads \citep{SasaoEtAl1980,DehantMathews2015}:
\begin{equation}
\bar{\omega}_\mathrm{fcn}=-1-\frac{\mathrm{A}}{\mathrm{A}_\mathrm{m}}(e_\mathrm{f}+K_\mathrm{cmb}-\beta)~.
\label{eq:omegafcnelasticKcmb}
\end{equation}
Importantly, $\bar{\omega}_\mathrm{cw}$ remains the same to this order of approximation, reflecting the fact this mode is only weakly affected by $\mathbf{\Gamma}_\mathrm{cmb}$ compared to the FCN. The solution Eqs.~\eqref{eq:mtwolayerelastic} \& \eqref{eq:mftwolayerelastic} remain valid, provided that one uses $\bar{\omega}_\mathrm{som}$ \& $\bar{\omega}_\mathrm{fcn}$ defined by Eqs.~\eqref{eq:SOMelasticKcmb} \& \eqref{eq:omegafcnelasticKcmb}.

\subsection{Transfer function}
\label{sec:transferfunctionKcmb}

\begin{figure}
   \centering
   \includegraphics[width=0.95\textwidth]{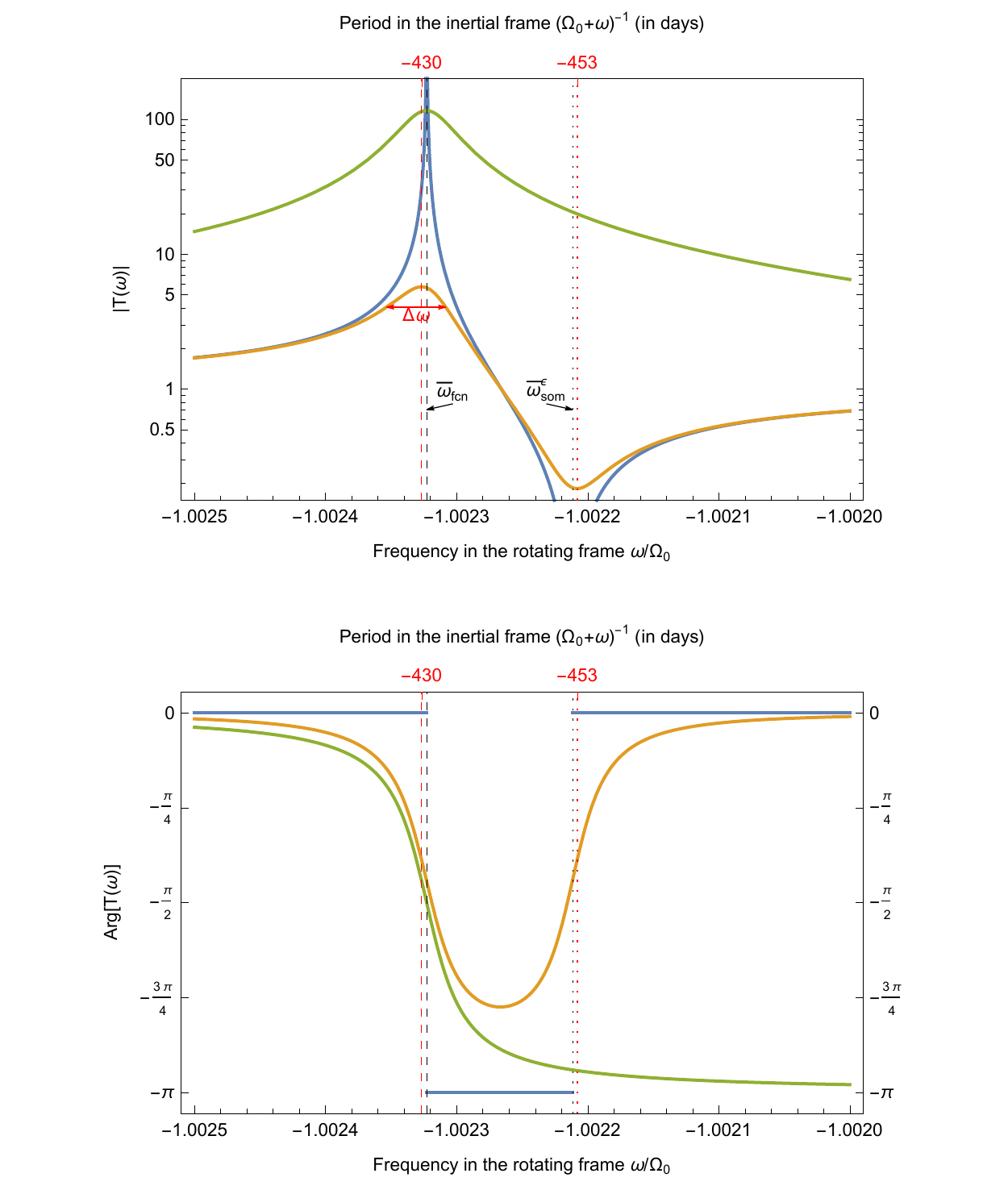} 
   \caption{Magnitude (top), and phase (bottom) of the transfer function for the two-layer elastic planet for the parameters of Table~\ref{tab:parameters}, with the exception that $e_\mathrm{f}+\mathrm{Re}(K_\mathrm{cmb})$ is treated as a constant (see main text). The blue curve corresponds to $K_\mathrm{cmb}=0$, the yellow one to $K_\mathrm{cmb}\neq0$, and the green one is similar to the yellow one but for $\bar{\omega}_\mathrm{som}^\epsilon\rightarrow-1$. {The black dashed and dotted lines indicate the positions of the frequencies, $\mathrm{Re}(\bar{\omega}_\mathrm{fcn})$ and $\mathrm{Re}(\bar{\omega}_\mathrm{som}^\epsilon)$ respectively. Their red equivalent respectively correspond to the maximum and minimum of the yellow curve.}}
   \label{fig:Bode_Earth}
\end{figure}

The fact that $K_\mathrm{cmb}$ is a complex number, introduces qualitative changes to the dynamics of the two-layer model besides the quantitative shifts in the values of $\bar{\omega}_\mathrm{som}$, and $\bar{\omega}_\mathrm{fcn}$. These can be appreciated by looking at the -- now complex -- transfer function, whose amplitude and phase are shown respectively on the top and bottom panels of Fig.~\ref{fig:Bode_Earth}. The yellow curve corresponds to the parameters of Table~\ref{tab:parameters}, and the blue curve is the same thing but for $K_\mathrm{cmb}=0$. To facilitate comparison, we have adjusted the value of $e_\mathrm{f}$ so that the numerical values of both $\bar{\omega}_\mathrm{fcn}$, and $\bar{\omega}_\mathrm{som}$ are the same in both curves. This amounts to treat the combination $e_\mathrm{f}+\mathrm{Re}(K_\mathrm{cmb})$ as a constant, which corresponds to what is actually measured in observations \citep[see][]{Dehant2017,ZhuEtAl2017}. In order to emphasise the role of $\bar{\omega}_\mathrm{som}^\epsilon$ we have also added the green curve, which is similar to the yellow one but with the substitution $\bar{\omega}_\mathrm{som}^\epsilon\rightarrow-1$  in Eq.~\eqref{eq:mtwolayerelastic}. This corresponds to the naive limit which we have already shown to be incompatible with the gyrostatic rigidity constraint in Sec.~\ref{sec:transferfunction}, and is here considered only for illustrative purposes.

There are two things to observe from the top panel. First, by comparing the yellow and green curves to the blue one, we see that having $\mathrm{Im}(K_\mathrm{cmb})\neq0$ causes the magnitude of the transfer function to remain finite for all values of the frequencies, even near the pole at $\bar{\omega}_\mathrm{fcn}$. Something similar can be observed near $\bar{\omega}_\mathrm{som}^\epsilon$ where the yellow curve never reaches zero, contrary to the blue one. This is the expected behaviour for dynamics systems described by a transfer function with complex zeros and/or poles \citep[see e.g.][]{FranklinEtAl1994}. Second, we can see that the maximum of the yellow curve is considerably smaller {and that it is slightly offset to the left compared to the green one, corresponding to a period of $\approx-1$ day in the inertial frame. As the two curves correspond to exactly the same coupling at the CMB, it is clear that this difference cannot by explained by the presence of dissipation alone.} {In the next section, we show that both of these observations can be interpreted as a consequence of} the proximity between the numerical values of $\bar{\omega}_\mathrm{fcn}$, and $\bar{\omega}_\mathrm{som}^\epsilon$. {This can already by appreciated} by looking at the bottom panel which shows that this proximity prevents a complete phase reversal of the yellow curve, contrary to the green. {Another more subtle effect of this proximity is to} slightly widen the {peak in the yellow} curve. This can be measured from the \textit{quality factor}, $Q$, defined as:
\begin{equation}
Q=\left|\frac{\omega_\mathrm{max}}{\Delta\omega}\right|~,
\end{equation}
where $\omega_\mathrm{max}$ is the frequency at resonance, and $\Delta\omega$ is the width of that resonance defined as the difference between the two values of $\omega$ for which $|\mathrm{T}(\omega)|=|\mathrm{T}(\omega_\mathrm{max})|/\sqrt{2}$. This last quantity is shown as the red double arrow on Fig.~\ref{fig:Bode_Earth}. For the yellow curve we find $Q_\mathrm{fcn}\approx22 000$, which is slightly lower than for the green curve $Q_\mathrm{fcn}\approx24 000$. 

\section{Discussion}
\label{sec:discussion}

{
\subsection{Offset in the FCN resonance frequency}
\label{sec:offset}
In the previous section, we have shown how the existence of dissipation at the CMB causes a slight offset of the maximum in the magnitude of the transfer function compared to the FCN frequency. This is a well-known phenomenon that is commonly described by analogy with the simpler system of a single damped harmonic oscillator. The response of this system to excitation can be represented by a transfer function with a single pole and no zero. If we call $\omega_0$, the free complex frequency of that system, one can then show that the damped resonance frequency, $\omega_d$, and quality factor are equal to \citep[][p. 122]{FranklinEtAl1994}:
\begin{align}
\omega_d=\mathrm{Re}(\omega_0)\sqrt{1-\frac{\mathrm{Im}(\omega_0)^2}{\mathrm{Re}(\omega_0)^2}}~,&&Q=\left|\frac{\mathrm{Re}(\omega_0)}{2\mathrm{Im}(\omega_0)}\right|~.
\label{eq:omegaQOH}
\end{align}
Applying the above to the Earth's, by setting $\omega_0=\bar{\omega}_\mathrm{fcn}$, and using the fact that $|\mathrm{Im}(\bar{\omega}_\mathrm{fcn})|\ll|\mathrm{Re}(\bar{\omega}_\mathrm{fcn})|\approx1$, we find:
\begin{equation}
\omega_\mathrm{max}\approx\mathrm{Re}(\bar{\omega}_\mathrm{fcn})-\frac{\mathrm{Im}(\bar{\omega}_\mathrm{fcn})^2}{2}~.\label{eq:omegafcnsimpledamped}
\end{equation}
This corresponds to a period shift of $\approx-3$ seconds, much smaller than that computed in Sec.~\ref{sec:transferfunctionKcmb}. For the quality factor, we find $Q\approx24000$. These two values are in agreement with those inferred from Fig.~\ref{fig:Bode_Earth} for the green curve, but not the yellow one. Since the difference between these two curves lies only in the value of $\omega_\mathrm{som}^\epsilon$ used, we deduce that the simple harmonic oscillator analogy works well only in the case when the values of $\omega_\mathrm{fcn}$ and $\omega_\mathrm{som}^\epsilon$ are not too close to each other. We can make this statement a little more precise if we define:
\begin{equation}
\delta\equiv\bar{\omega}_\mathrm{fcn}-\bar{\omega}_\mathrm{som}^\epsilon~.
\label{eq:deltadef}
\end{equation}
One can then show that the magnitude of the transfer function is maximum at:
\begin{equation}
\omega_\mathrm{max}=\mathrm{Re}(\bar{\omega}_\mathrm{fcn})-\frac{\mathrm{Re}(\delta)}{2}\left(1+\frac{\mathrm{Im}(\delta)^2}{\mathrm{Re}(\delta)^2}-\frac{\mathrm{Im}(\delta)\mathrm{Im}(\bar{\omega}_\mathrm{fcn})}{\mathrm{Re}(\delta)^2}\right)+
\frac{|\delta|}{2}\sqrt{1+\left(\frac{\mathrm{Im}(\delta)-2\mathrm{Im}(\bar{\omega}_\mathrm{fcn})}{\mathrm{Re}(\delta)}\right)^2}~.\label{eq:omegamax}
\end{equation}
The same function is minimum at the frequency:
\begin{equation}
\omega_\mathrm{min}=\mathrm{Re}(\bar{\omega}_\mathrm{fcn})-\frac{\mathrm{Re}(\delta)}{2}\left(1+\frac{\mathrm{Im}(\delta)^2}{\mathrm{Re}(\delta)^2}-\frac{\mathrm{Im}(\delta)\mathrm{Im}(\bar{\omega}_\mathrm{fcn})}{\mathrm{Re}(\delta)^2}\right)-
\frac{|\delta|}{2}\sqrt{1+\left(\frac{\mathrm{Im}(\delta)-2\mathrm{Im}(\bar{\omega}_\mathrm{fcn})}{\mathrm{Re}(\delta)}\right)^2}~.\label{eq:omegamin}
\end{equation}
These expressions are particularly enlightening in the limit $\delta\rightarrow0$, where they converge to:
\begin{align}
\omega_\mathrm{max}&\rightarrow\mathrm{Re}(\bar{\omega}_\mathrm{fcn})-|\mathrm{Im}(\bar{\omega}_\mathrm{fcn})|~,\label{eq:limitomegamax}\\
\omega_\mathrm{min}&\rightarrow\mathrm{Re}(\bar{\omega}_\mathrm{fcn})+|\mathrm{Im}(\bar{\omega}_\mathrm{fcn})|~.\label{eq:limitomegamin}
\end{align}
Eq.~\eqref{eq:limitomegamax}, shows that when the FCN and SOM frequencies are very close, the offset produced on the resonance frequency is linear in the quantity $|\mathrm{Im}(\bar{\omega}_\mathrm{fcn})|$ representing the mode's damping. This is in contrast with the result of Eq.~\eqref{eq:omegafcnsimpledamped} for the simple harmonics oscillator where this offset is quadratic. 
The fact that the simple model fails to predict the correct resonance frequency should not be a surprise given the superior complexity of the two-layer nutation model with dissipation. What the above computation shows is that this difference in behaviour can be interpreted mathematically as the result of having a complex zero at a frequency close to that of the FCN in the system's transfer function. Similar behaviours can be found in other physical systems described by transfer functions \citep[see again][]{FranklinEtAl1994}. The fact that in the present case, the zero's frequency corresponds to $\bar{\omega}_\mathrm{som}^\epsilon$ is also interesting.} 

Despite their proven limitations, Eqs.~\eqref{eq:omegaQOH} are commonly used in the literature to identify the frequency and quality factor of the FCN from the observations of its resonance with the nutations \citep[e.g.][]{koot2010,ChaoHsieh2015}. In the following subsection, we reformulate the main elements of the above in the standard formalism in order to facilitate comparison with other works. 

\subsection{Comparison to previous work}
\label{sec:comparison}
In order to compare observations to theory, \citet{MathewsEtAl2002} introduced the following parametrisation of the transfer function of an elastic planet {with dissipation}:
\begin{equation}
\mathrm{T}(\omega)=\frac{\omega_\mathrm{E}-\omega}{1+\omega_\mathrm{E}}\left(1+(1+\omega)\sum_{\alpha=1}\frac{N_\alpha}{\omega-\omega_\alpha}\right)~,
\label{eq:TMHB}
\end{equation}
where the sum runs over the set of free modes with frequencies, $\omega_\alpha$. There are two such modes in the two-layer system: the FCN, and the CW. For a three-layer planet with a solid inner core, such as the Earth, one must add to this list the \textit{Free Inner Core Nutation} (FICN), and the \textit{Inner Core Wobble} (ICW). However, these two modes only have a small influence on the nutations compared to the FCN, and CW, and may therefore be excluded from the analysis in first approximation. The transfer function Eq.~\eqref{eq:TMHB} satisfies the gyrostatic rigidity constraint: $\mathrm{T}(-1)=1$, as well as $\mathrm{T}(\omega_\mathrm{E})=0$. We can compute the expressions of the coefficients $N_\alpha$ analytically by direct comparison to the (normalised) transfer function of previous sections:
\begin{align}
N_\mathrm{cw}&=-\frac{\mathrm{A}}{\mathrm{A}_\mathrm{m}}\left(1-\frac{\kappa}{e}+\frac{\epsilon-\kappa}{e}\bar{\omega}_\mathrm{cw}\right)\frac{\left(\bar{\omega}_\mathrm{cw}-\bar{\omega}_\mathrm{som}^\epsilon\right)}{\left(\bar{\omega}_\mathrm{cw}-\bar{\omega}_\mathrm{fcn}\right)}~,\label{eq:Ncw}\\
N_\mathrm{fcn}&=-\frac{\mathrm{A}}{\mathrm{A}_\mathrm{m}}\left(1-\frac{\kappa}{e}+\frac{\epsilon-\kappa}{e}\bar{\omega}_\mathrm{fcn}\right)\frac{\left(\bar{\omega}_\mathrm{fcn}-\bar{\omega}_\mathrm{som}^\epsilon\right)}{\left(\bar{\omega}_\mathrm{cw}-\bar{\omega}_\mathrm{fcn}\right)}\frac{(1+\bar{\omega}_\mathrm{cw})}{(1+\bar{\omega}_\mathrm{fcn})}~.\label{eq:Nfcn}
\end{align}
To leading order in the small quantities, Eqs.~\eqref{eq:Ncw} \& \eqref{eq:Nfcn} reduce to:
\begin{align}
N_\mathrm{cw}&\approx-\frac{\mathrm{A}}{\mathrm{A}_\mathrm{m}}\left(1-\frac{\kappa}{e}\right)=-\frac{\bar{\omega}_\mathrm{cw}}{e}~,\label{eq:Ncw1}\\
N_\mathrm{fcn}&\approx\frac{\mathrm{A}_\mathrm{f}}{\mathrm{A}_\mathrm{m}}\left(1-\frac{\gamma}{e}\right)~,\label{eq:Nfcn1}
\end{align}
in agreement with \citet{MathewsEtAl2002}. As stated by these authors, direct measurements of the strengths, and locations of the resonances in the amplitude of the nutations can then be used to constrain the parameters that appear in the explicit expressions for $\bar{\omega}_\mathrm{fcn}$, $\bar{\omega}_\mathrm{cw}$, and $N_\mathrm{fcn}$. The transfer function obtained from the values of Eqs.~\eqref{eq:Ncw1} \& \eqref{eq:Nfcn1} is undistinguishable to that of Fig.~\ref{fig:Bode_Earth} (yellow curve). However, {as we have shown in the previous subsection,} caution must be exerted in identifying the {frequency of the resonance with the FCN}. 

{
Using Eq.~\eqref{eq:Nfcn1}, we can rewrite Eq.~\eqref{eq:deltadef} as: 
\begin{align}
\delta&=\left(-\frac{\mathrm{A}_\mathrm{f}}{\mathrm{A}_\mathrm{m}}+\frac{\epsilon}{e-\epsilon}\right)\left(e_\mathrm{f}-\beta+K_\mathrm{cmb}\right)\label{eq:deltaexpl}\\
&\approx\frac{N_\mathrm{fcn}}{1+N_\mathrm{fcn}}\left(1+\bar{\omega}_\mathrm{fcn}\right)~.
\label{eq:deltaNfcn}
\end{align}
From Eq.~\eqref{eq:Nfcn} we can see clearly that, in addition to causing the frequency offset computed in the previous section, having $\delta$ small also causes the decrease in the amplitude of the resonance. This is another expected effect of having a pole close to a zero in the transfer function \citep[see again][p. 131]{FranklinEtAl1994}. This effect can be clearly observed from the comparison between the green and the yellow curves on Fig.~\ref{fig:Bode_Earth}. The fact that $N_\mathrm{fcn}\rightarrow0$ when $\delta\rightarrow0$ follows immediately from Eq.~\eqref{eq:Nfcn}, but it is not so obvious from the approximate expression Eq.~\eqref{eq:Nfcn1}. It becomes clearer when one realises that this limit corresponds to having $\mathrm{A}_\mathrm{f}\ll\mathrm{A}_\mathrm{m}$ in Eq.~\eqref{eq:deltaexpl}.}

Using the above, the transfer function derived in Sec.~\ref{sec:transferfunctionKcmb} can be approximated by:
\begin{align}
\mathrm{T}(\omega)&\approx \frac{\mathrm{A}}{\mathrm{A}_\mathrm{m}}\frac{(e-\epsilon)}{e}\frac{\left(\omega-\bar{\omega}_\mathrm{som}^\epsilon\right)}{\left(\omega-\bar{\omega}_\mathrm{fcn}\right)}\\
&\approx1+\frac{N_\mathrm{fcn}(1+\omega)}{(\omega-\bar{\omega}_\mathrm{fcn})}~.\label{eq:transferfunctiondiurnalNfcn}
\end{align}
which is valid in the near-diurnal band where $\omega\approx-1$. Incidentally, note that Eq.~\eqref{eq:transferfunctiondiurnalNfcn} satisfies the gyrostatic rigidity constraint exactly, and that we could have arrived to it directly from Eq.~\eqref{eq:TMHB}. 

\begin{figure}
   \centering
   \includegraphics[width=0.80\textwidth]{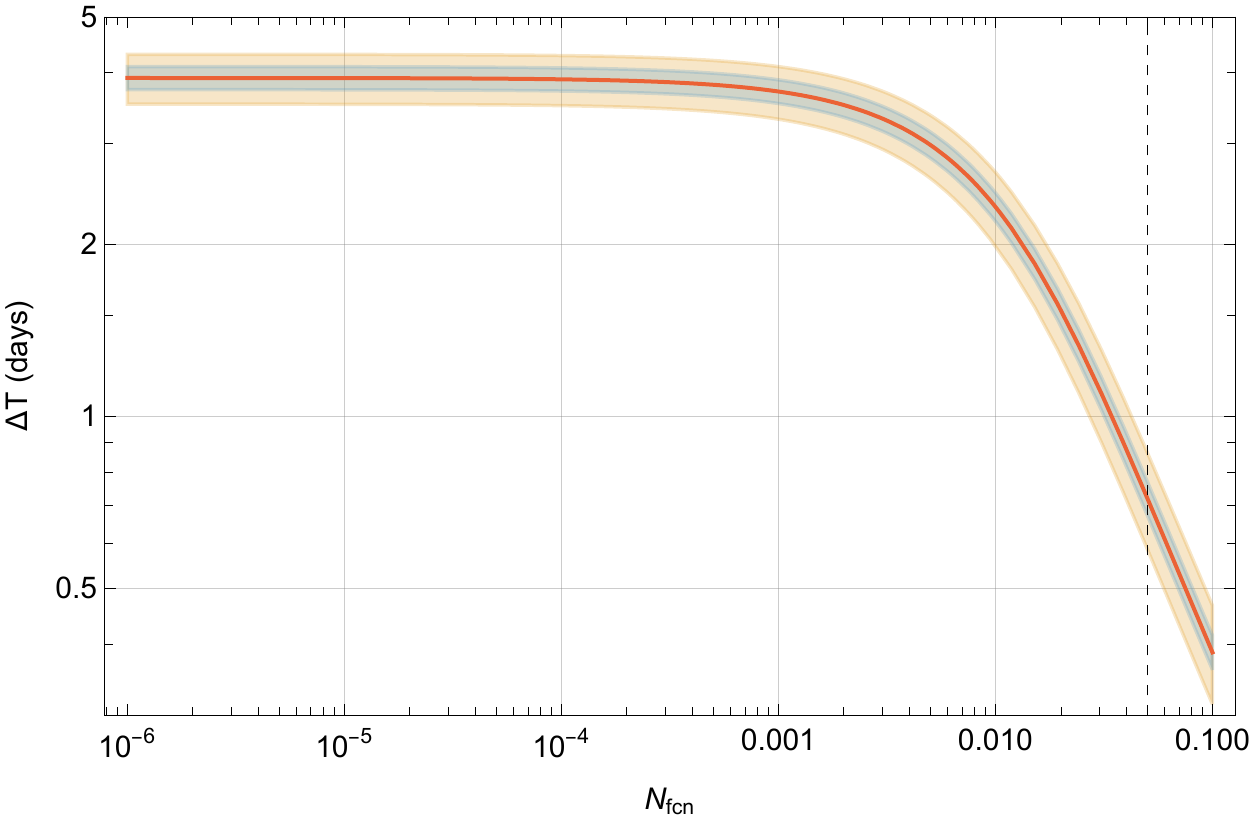} 
   \caption{Period offset in the inertial frame, $\Delta T$, as given by Eq.~\eqref{eq:DeltaT} as a function of $N_\mathrm{fcn}$ (other parameters are from Table~\ref{tab:parameters}). The inner shaded area (in blue) corresponds to the interval of $\pm10$ days about $T_\mathrm{fcn}$, and the outer one (in yellow) to the interval of $\pm10\%$ about $\mathrm{Im}(K_\mathrm{cmb})$. The vertical dashed line corresponds to the Earth's value.}
   \label{fig:DeltaT}
\end{figure}

From Eq.~\eqref{eq:omegamax}, we can compute the \textit{period offset}, $\Delta T$, in the inertial frame given by the formula:
\begin{align}
\Delta T&\equiv T_\mathrm{max}-T_\mathrm{fcn}\nonumber\\
&=\frac{\mathrm{Re}(\bar{\omega}_\mathrm{fcn})-\omega_\mathrm{max}}{(1+\mathrm{Re}(\bar{\omega}_\mathrm{fcn}))(1+\omega_\mathrm{max})}~.\label{eq:DeltaT}
\end{align}
The result is shown in red on Fig.~\ref{fig:DeltaT} as a function of $N_\mathrm{fcn}$, for the parameters of Table~\ref{tab:parameters}.
We see that $\Delta T$ gets larger for decreasing values of $N_\mathrm{fcn}$ before reaching a plateau corresponding to the limit of Eq.~\eqref{eq:limitomegamax}. From Eq.~\eqref{eq:DeltaT}, we see that the level of this plateau depends on the values of both $T_\mathrm{fcn}$, and $\mathrm{Im}(K_\mathrm{cmb})$, respectively through $\mathrm{Re}(\bar{\omega}_\mathrm{fcn})$, and $\mathrm{Im}(\bar{\omega}_\mathrm{fcn})$. The inner shaded area of the figure (in blue) corresponds to the interval of $\pm10$days about the value of $T_\mathrm{fcn}$ derived from Table~\ref{tab:parameters}. Similarly, the outer shaded area (in yellow) corresponds to the interval of $\pm10\%$ about $\mathrm{Im}(K_\mathrm{cmb})$. Those deviations are far bigger than current uncertainties in measurements, and are here provided as an indication of the weak dependence of $\Delta T$ on these quantities. The vertical dashed line corresponds to the best-fit value of $N_\mathrm{fcn}$ for the Earth.

\subsection{Implications for Mars and the Moon}

Closer values of $\bar{\omega}_\mathrm{fcn}$, and $\bar{\omega}_\mathrm{som}^\epsilon$, lead to a larger offset, $\Delta T$, between the resonance period, $T_\mathrm{max}$, and $T_\mathrm{fcn}$. From Eq.~\eqref{eq:deltaNfcn}, we therefore expect this offset to be larger in situations where $N_\mathrm{fcn}\ll1$, and $T_\mathrm{fcn}\rightarrow\infty$. From Eq.~\eqref{eq:Nfcn1}, we see that this first condition corresponds to $\mathrm{A}_\mathrm{f}/\mathrm{A}_\mathrm{m}\ll1$. For Mars, this ratio is expected to be roughly of the same order of magnitude as in the Earth, and $T_\mathrm{fcn}^\mathrm{\mars}\approx-240$ days \citep{FolknerEtAl1997,VanHoolstEtAl2000}. Furthermore, since Mars has no intrinsic global magnetic field, the dissipation at the CMB is mainly viscous, and so probably much smaller than in the Earth. Therefore, we expect $\Delta T$ to be negligible for this planet, {to the effect that the observed resonance period can be safely equated to $T_\mathrm{fcn}$ and used as such to constrain the basic rotation parameters}.

The situation is different for the Moon, where Lunar Laser Ranging (LLR) measurements give $\mathrm{A}_\mathrm{f}/\mathrm{A}\approx7\times10^{-4}$, $e_\mathrm{f}\approx 2.2\times10^{-4}$, and $\mathrm{Im}(K_\mathrm{cmb})\approx-2.34\times10^{-5}$ \citep{WilliamsEtAl2014}, the latter value being computed from the formula given by \citet{OrganowskiDumberry2020}. {Note that the flow at the Moon's CMB is likely turbulent \citep{WilliamsEtAl2001}, so that this latter parameter is proportional to $|\mathrm{m}_\mathrm{f}|$ \citep{Yoder1981,CebronEtAl2019}.} Since the Moon's spin rate is much slower than the Earth's, it is more useful to give the ratio: $\Delta T/T_\mathrm{fcn}$, which in the limit of $N_\mathrm{fcn}$ and $\mathrm{Im}(K_\mathrm{cmb})$ small approximates to:
\begin{equation}
\frac{\Delta T}{T_\mathrm{fcn}}\rightarrow-\frac{|\mathrm{Im}(K_\mathrm{cmb})|}{e_\mathrm{f}}~,
\label{eq:DT/Tfcn}
\end{equation}
which is about $-10\%$ for the Moon. Equation \eqref{eq:DT/Tfcn} could, in principle, be used as an independent means to measure $\mathrm{Im}(K_\mathrm{cmb})$. Something that might however prove difficult, as the FCN period is both very long, and poorly constrained at the moment: $T_\mathrm{fcn}^\mathrm{\leftmoon}\approx-374\pm93$ years \citep{ViswanathanEtAl2019}. Also, as we have already noted, having $N_\mathrm{fcn}\ll1$, considerably reduces the amplitude of the resonance. Indirect observation of the FCN based on its resonance with weaker modes of physical libration have been proposed \citep{BarkinEtAl2014,PetrovaEtAl2018}. Our results show that such measurements would not give $T_\mathrm{fcn}^\mathrm{\leftmoon}$ directly, but rather a value $10\%$ smaller. A possible way to obtain this mode's frequency directly, would be to combine measurements of $\omega_\mathrm{max}$, and $\omega_\mathrm{min}$ with Eqs.~\eqref{eq:limitomegamax} \& \eqref{eq:limitomegamin}.

\subsection{Implications for the Earth's secular evolution}

Throughout this work, we have treated the parameters of Table~\ref{tab:parameters} as constants. While this is certainly true on the timescales relevant for geodetic observation, these parameters might have changed significantly throughout the Earth's evolution. \citet{Greff-LefftzLegros1999,Greff-LefftzLegros1999a} examined modulations in the FCN frequency caused by changes in the ratio of moments of inertia, and in the CMB flattening over geological timescales, as well as the secular deceleration of the axial spin rate, $\Omega_0$ caused by the lunar tidal torque. They identified the latter as the dominant source of FCN modulation which could have caused episodes of increased dissipation at the CMB by `tuning' the FCN into resonance with the luni-solar tidal waves. They argued that such episode might be correlated to major geological events. They also estimate that secular variations of $K_\mathrm{cmb}$ could not have significantly influenced this tuning mechanism, given the weak dependence of $\mathrm{Re}(\omega_\mathrm{fcn})$, on this parameter. It might be worthwhile revisiting that argument in light of the present study, especially given the strong dependence of the frequency offset on the value of $\mathrm{Im}(\mathrm{K}_\mathrm{cmb})$.

\subsection{Conclusion}

{In the present work, we have looked in details at the differences between the SOM and the FCN. We elaborated on the fact that, despite their similarity, those two motions are not the same thing, and that they are free modes of two different systems, namely that of a fluid core inside a steadily-rotating mantle, and that of the whole planet when the mantle can wobble freely. Focusing on that later system subjected to an external tidal forcing, we have shown that the frequency of the SOM retains some significance in the non-dissipative freely-rotating model and corresponds to that where the external torque on the mantle is exactly balanced by that produced by the core at the CMB, thereby leaving the free mantle in a state of steady rotation. When dissipation is reintroduced, we have shown that the peak of the resonance of the forced nutations with the FCN is slightly offset by a value of $\approx-1$ day
for the Earth. We explained how this can be interpreted mathematically as the result of the presence of a complex zero of the transfer function at the frequency of the SOM close to the pole at the FCN frequency. We demonstrated how the offset is larger when these two frequencies are very close to each other. Based on that, we showed that this offset is likely to be negligible for Mars, but could be more important for the Moon ($\approx-10\%$ of the FCN frequency). We also briefly discussed how this result might have been different in other periods during the Earth's evolution.}

{
In future work, it might be interesting to reconsider the above formalism based on the transfer function, and how it may relate to the work of \citet{Triana2019} who showed that the FCN may interact in a complex way with other inertial modes present in the viscous core of the Earth when their frequencies are close. Something that may help cast light on some of the remaining elusive properties of inertial modes coupled to planetary rotation.
}

\begin{acknowledgments}
The author would like to thank Profs. V. Dehant, T. Van Hoolst, and Dr. S. A. Triana for the very useful discussions throughout the writing of this work. He is also thankful to the two reviewers Profs. B. Buffett and M. Dumberry whose comments and careful reviews helped to significantly improve the manuscript.
J.R. acknowledges funding by the European Research Council under the European Union’s Horizon 2020 research and innovation program (\textsc{graceful}, Synergy Grant agreement No 855677).
\end{acknowledgments}

\bibliography{MyLibrary}
\bibliographystyle{aasjournal}


\begin{appendix}

\section{Transformation between inertial and rotating frames}
\label{sec:frames}
We can parametrise the \textit{passive} transformation relating the components of a vector, $\mathbf{v}$, in the inertial frame coordinates, $\{V^x,V^y,V^z\}$, to the components of the same vectors in the body frame coordinates, $\{v^x,v^y,v^z\}$ as:
\begin{equation}
\left(
\begin{array}{c}
 v^x\\
 v^y\\
 v^z\\ 
\end{array}
\right)
=\underbrace{R_3(-\gamma)\cdot R_2(-\beta)\cdot R_1(-\alpha)\cdot R_3(-\Omega_0 t)}_{T}
\left(
\begin{array}{c}
 V^x \\
 V^y \\
 V^z 
\end{array}
\right)~.
\label{eq:in2rot}
\end{equation}
where $\{R_i(\theta)\}$ are the set of the usual three-dimensional matrix representation of rotation by an angle, $\theta$:
\begin{align}
R_1(\theta)=\left(
\begin{array}{ccc}
 1 & 0 & 0 \\
 0 & \cos\theta & -\sin\theta \\
 0 & \sin\theta & \cos\theta \\
\end{array}
\right),&&
R_2(\theta)=\left(
\begin{array}{ccc}
 \cos\theta& 0 & -\sin\theta\\
 0 & 1 & 0\\  
 \sin\theta& 0 & \cos\theta
\end{array}
\right),&&
R_3(\theta)=\left(
\begin{array}{ccc}
 \cos\theta & -\sin\theta & 0\\
 \sin\theta & \cos\theta & 0\\
  0 & 0 & 1\\
\end{array}
\right)~.
\end{align}
The three Euler angles, $\alpha$, $\beta$, $\gamma$, relate the coordinates in the body frame to those in the frame rotating steadily at the diurnal frequency (SRF, see above). They are related to the components of the angular velocity, $\mathbf{\Omega}$, via dynamical relations and must therefore be treated as first order quantities in agreement with the limit $|\mathbf{m}|\ll1$. Based on Eq.~\eqref{eq:in2rot}, we can immediately compute the components of the vector $\hat{\mathbf{Z}}$ used in Eq.~\eqref{eq:OmegaSRF} of Sec.~\ref{sec:eqmotion} in the body frame to be $(-\beta,\alpha,1)^\intercal$. From the time-dependent transformation matrix, $T$, in Eq.~\eqref{eq:in2rot}, we can define the rotation tensor:
\begin{equation}
\underline{\mathbf{R}}\equiv \frac{dT}{dt}\cdot T^\intercal~.
\end{equation}
This tensor is antisymmetric. Its dual vector defines the angular velocity of the transformation $T$, which must be equal to $\mathbf{\Omega}$, by definition \citep[see e.g.][for details]{Arnold1989}. Solving for $\alpha$, $\beta$, and $\gamma$ gives (in Fourier space):
\begin{align}
\alpha=-\frac{(i \omega\mathrm{m}^x+\Omega_0\mathrm{m}^y)\Omega_0}{\omega^2-\Omega_0^2}~,&&\beta=\frac{(-i \omega\mathrm{m}^y+\Omega_0\mathrm{m}^x)\Omega_0}{\omega^2-\Omega_0^2}~,&&\gamma=0~.
\label{eq:Euleranglesolution}
\end{align}
Upon reintroducing into Eq.~\eqref{eq:in2rot}, we find:
\begin{equation}
\tilde{v}=\tilde{V}\mathrm{e}^{-i\Omega_0 t}~,
\label{eq:vtildeVtilde}
\end{equation}
where we have defined, $\tilde{v}=(v^x+iv^y)$, and $\tilde{V}=(V^x+iV^y)$ as usual. From Eq.~\eqref{eq:vtildeVtilde}, we see that there is a diurnal phase difference between the components of any vector, $\mathbf{v}$, measured in the rotating frame and the same components measured in the inertial frame. Note that this result is consistent with the first component of Eq.~\eqref{eq:m-m+(t)}, the latter representing an \textit{active} rotation of a vector in a fixed system of coordinates.

The nutation vector has cartesian components in the inertial frame given by $(X,Y,0)^\intercal$, where $X$, and $Y$ denote the equatorial coordinates of the Earth's North pole with respect to the Celestial Reference Frame of J2000 \citep[see][]{DehantMathews2015}. By definition, the rate of change of this vector must be equal to $\mathbf{\Omega}\times\mathbf{r}$. Using Eq.~\eqref{eq:vtildeVtilde}, we therefore arrive at:
\begin{equation}
\Omega_0\tilde{\mathrm{m}}=i\frac{d\tilde{\eta}}{dt}\mathrm{e}^{-i\Omega_0 t}~,
\end{equation}
where we have defined $\tilde{\eta}=X+iY$. One must exert caution when computing the time derivative, $d\tilde{\eta}/dt$. In Fourier space we have: 
\begin{align}
\Omega_0\tilde{\mathrm{m}}(\omega)\mathrm{e}^{i\omega t}&=i\frac{d}{dt}\left(\tilde{\eta}(\omega)\mathrm{e}^{i\omega t}\right)\mathrm{e}^{-i\Omega_0 t}\\
&=-\omega\tilde{\eta}(\omega)\mathrm{e}^{i(\omega-\Omega_0)t}~,
\end{align}
from which we see that the individual Fourier component of frequency $\omega$ in the wobble, $\tilde{\mathrm{m}}(\omega)$, is proportional to the individual Fourier component of frequency $(\omega+\Omega_0)$ in the nutation, $\tilde{\eta}(\omega+\Omega_0)$. \textit{In extenso:}
\begin{equation}
\tilde{\eta}(\omega+\Omega_0)=-\frac{\tilde{\mathrm{m}}(\omega)}{\omega+\Omega_0}\Omega_0~.
\label{eq:wobblenutation}
\end{equation}
This expression becomes infinite for $\omega=-\Omega_0$, corresponding to the resonance with the TOM (see Sec.~\ref{sec:SOM}).

\section{Spin-Over solution from the momentum equation of fluid dynamics}
\label{sec:fluiddyn}
The momentum equation of a fluid in the frame rotating at angular velocity, $\mathbf{\Omega}$, given by Eq.~\eqref{eq:Omegam} writes, to first order in the velocity, and in the Fourier space:
\begin{equation}
i\omega(\mathbf{v}+\mathbf{m}\times\mathbf{r})+2\mathbf{\Omega}\times\mathbf{v}=-\mathbf{\nabla}p~,
\label{eq:momentum}
\end{equation}
where $p$, and $\mathbf{v}$ respectively denote the (reduced) pressure and the velocity of the fluid. The latter can be written using the ansatz Eq.~\eqref{eq:v=wxr+gradpsi}. The scalar function, $\psi$, is then chosen to accommodate the `no-penetration' boundary condition at the CMB:
\begin{align}
\mathbf{v}\cdot\hat{\mathbf{n}}|_{\partial\mathcal{V}}&=0\\
\leftrightarrow\hat{\mathbf{n}}\cdot\mathbf{\nabla}\psi|_{\partial\mathcal{V}}&=-\hat{\mathbf{n}}\cdot(\boldsymbol{\omega}\times\mathbf{r})|_{\partial\mathcal{V}}~,
\end{align}
where $\partial\mathcal{V}$ is the surface boundary of the fluid volume, $\mathcal{V}$, with normal vector $\hat{\mathbf{n}}$. The incompressibility condition, $\mathbf{\nabla}\cdot\mathbf{v}=0$, imposes that $\psi$ must be an harmonic function of the coordinates, i.e. $\nabla^2\psi=0$. \citet{Poincare1910} computed this function for the fluid ellipsoid with dimensions $(a,b,c)$. In the special case where $b=a$, using Eq.~\eqref{eq:omegaf}, it writes:
\begin{align}
\psi&=\Omega_0\frac{a^2-c^2}{a^2+c^2}(\mathrm{m}_\mathrm{f}^y x-\mathrm{m}_\mathrm{f}^x y)z\\
&=\Omega_0e_\mathrm{f}(\mathrm{m}_\mathrm{f}^y x-\mathrm{m}_\mathrm{f}^x y)z~.
\label{eq:psi}
\end{align}
where we have used the definition $e_\mathrm{f}=(\mathrm{C}_\mathrm{f}-\mathrm{A}_\mathrm{f})/\mathrm{A}_\mathrm{f}$, where $\mathrm{A}_\mathrm{f}$, and $\mathrm{C}_\mathrm{f}$ are the equatorial, and polar moments of inertia of the axisymmetric fluid ellipsoid:
\begin{align}
\mathrm{A}_\mathrm{f}=\mathrm{M}_\mathrm{f}\frac{\left(a^{2}+c^{2}\right)}{5}~,&&
\mathrm{C}_\mathrm{f}=2\mathrm{M}_\mathrm{f}\frac{a^{2}}{5}~,
\end{align}
where, $\mathrm{M}_\mathrm{f}=(4\pi/3)\rho_\mathrm{f}a^2c$, is the ellipsoid's mass, and $\rho_\mathrm{f}$, its density.

Taking the curl of Eq.~\eqref{eq:momentum}, we have, to first order in $\mathbf{m}$ and $\mathbf{m}_\mathrm{f}$:
\begin{equation}
i\omega(\mathbf{m}+\mathbf{m}_\mathrm{f})=\hat{\mathbf{z}}\times\mathbf{m}_\mathrm{f}+\Omega_0^{-1}\left(\hat{\mathbf{z}}\cdot\mathbf{\nabla}\right)\mathbf{\nabla}\psi~.
\label{eq:curlmomentum}
\end{equation}
Plugging in Eq.~\eqref{eq:psi}, the equatorial components of Eq.~\eqref{eq:curlmomentum} can be combined into:
\begin{equation}
\tilde{\mathrm{m}}\omega=-\tilde{\mathrm{m}}_\mathrm{f}\left(\omega+(1+e_\mathrm{f})\Omega_0\right)~,
\end{equation}
which, upon solving for $\tilde{\mathrm{m}}_\mathrm{f}$, gives us Eq.~\eqref{eq:mforcedmf}. If we replace $e_\mathrm{f}$ by its expression in terms of the ellipsoid's dimensions, we recover the familiar expression for the SOM frequency of fluid dynamics \citep{Greenspan1968}:
\begin{equation}
\omega_\mathrm{som}=-\frac{2a^2}{a^2+c^2}\Omega_0~.
\end{equation}
In this form, we see that the SOM frequency is proportional to the ratio of the polar and equatorial moments of inertia.

\end{appendix}

\end{document}